\documentclass[lettersize,journal]{IEEEtran}
\usepackage{amsmath,amsfonts}
\usepackage{algorithmic}
\usepackage{algorithm}
\usepackage{array}
\usepackage{textcomp}
\usepackage{stfloats}
\usepackage{url}
\usepackage{verbatim}
\usepackage{cite}
\usepackage[utf8]{inputenc} 
\usepackage[T1]{fontenc}    
\usepackage{hyperref}       
\usepackage{url}            
\usepackage{booktabs}       
\usepackage{amsfonts}       
\usepackage{nicefrac}       
\usepackage{microtype}      
\usepackage{xcolor}         
\usepackage{graphicx}
\usepackage{subcaption}

\usepackage{amsmath}
\usepackage{amsthm}

\usepackage{wrapfig}
\usepackage{multirow}
\usepackage{multicol}
\usepackage{enumitem}

\usepackage{pifont}
\usepackage{color, colortbl}

\usepackage{arydshln} 
\usepackage{booktabs}

\newcommand{\para}[1]{\vspace{2mm} \noindent \textbf{#1}}

\begin{document}

\title{Empowering Heterogeneous Graph Foundation Models via Decoupled Relation Alignment}

\author{Ziyu Zheng, Yaming Yang, Zhe Wang, Ziyu Guan$^{*}$, Wei Zhao
\thanks{* Corresponding author}
\thanks{Z. Zheng, Y. Yang, Z. Wang, Z. Guan, and W. Zhao are with the School of Computer Science and Technology, Xidian University, Xi'an, China 710071. E-mail: \{zhengziyu@stu., zwang\_01@stu., yym@,  zyguan@, ywzhao@mail.\}xidian.edu.cn}%
}



\maketitle

\begin{abstract}
While Graph Foundation Models (GFMs) have achieved remarkable success in homogeneous graphs, extending them to multi-domain heterogeneous graphs (MDHGs) remains a formidable challenge due to cross-type feature shifts and intra-domain relation gaps. Existing global feature alignment methods (PCA or SVD) enforce a shared feature space blindly, which distorts type-specific semantics and disrupts original topologies, inevitably leading to "Type Collapse" and "Relation Confusion". To address these fundamental limitations, we propose Decoupled relation Subspace Alignment (DRSA), a novel, plug-and-play relation-driven alignment framework. DRSA fundamentally shifts the paradigm by decoupling feature semantics from relation structures. Specifically, it introduces a dual-relation subspace projection mechanism to coordinate cross-type interactions within a shared low-rank relation subspace explicitly. Furthermore, a feature-structure decoupled representation is designed to decompose aligned features into a semantic projection component and a structural residual term, adaptively absorbing intra-domain variations. Optimized via a stable alternating minimization strategy based on Block Coordinate Descent, DRSA constructs a well-calibrated, structure-aware latent space. Extensive experiments on multiple real-world benchmark datasets demonstrate that DRSA can be seamlessly integrated as a universal preprocessing module, significantly and consistently enhancing the cross-domain and few-shot knowledge transfer capabilities of state-of-the-art GFMs. The code is available at: \url{https://github.com/zhengziyu77/DSRA}.
\end{abstract}

\section{Introduction}
\IEEEPARstart{G}{raph} data, as a powerful paradigm for modeling complex entity relationships, has been extensively adopted across diverse domains, including academic networks~\cite{ie-hgcn}, recommendation systems~\cite{rec, Trustrecom}, and bioinformatics analysis~\cite{mol}. In recent years, inspired by the success of large-scale pre-training in the language~\cite{nlpfd1} and vision fields~\cite{cvfd1}, Graph Foundation Models (GFMs)~\cite{gfm, gfm2} have emerged as a pivotal direction in graph representation learning. By pre-training on multi-source graph data and adapting to downstream tasks, GFMs aim to transfer generalizable knowledge across different scenarios. Recent studies~\cite{samgpt,gcope,mdgfm,graver,RAG-GFM} indicate that joint modeling over multi-domain graphs facilitates the acquisition of universal structural patterns and semantic representations, thereby improving generalization under distribution shifts~\cite{advdomain}.

Despite remarkable progress, current research on multi-domain GFMs primarily focuses on homogeneous graphs~\cite{samgpt,gcope,mdgfm,graver,RAG-GFM, graphprompt,msgcot,gpf}, which assume uniform node and edge types and merely require addressing inter-domain distribution shifts. In contrast, extending this paradigm to multi-domain heterogeneous graphs remains a formidable challenge. Heterogeneous graphs exhibit not only cross-domain distribution shifts but also complex intra-domain structures composed of multi-type nodes and diverse relation semantics~\cite{shgp, rash,han,heco,hero}, which significantly exacerbates the difficulty of cross-domain knowledge transfer. 
\begin{figure}[t]
\centering
\includegraphics[width=0.95\linewidth]{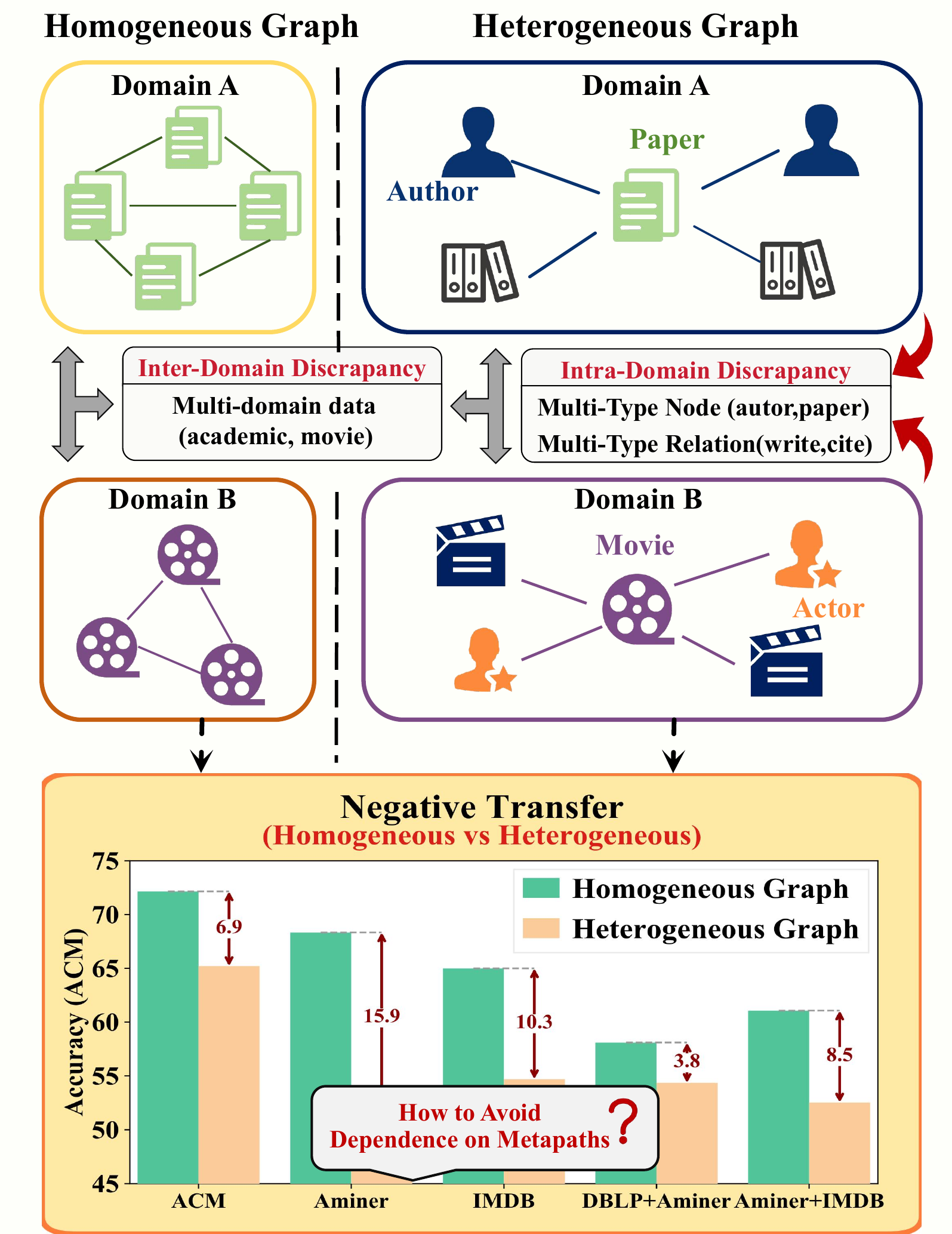}
\caption{Multi-domain heterogeneous graph foundation models exhibit significantly different negative transfer behaviors from the perspectives of meta-path-based homogeneous graphs and raw heterogeneous relation graphs.}
\label{motivation1}
\end{figure}
To investigate the impact of this dual-level discrepancy, we conduct multi-domain pre-training using the ACM dataset~\cite{heco} as the target domain. Employing the same pre-training strategy on the source domains, we utilize meta-path graphs and raw heterogeneous graphs as inputs, respectively. Experimental results reveal a pronounced negative transfer phenomenon from both perspectives, which becomes even more severe when relation aggregation is directly performed on the raw heterogeneous graphs. This finding indicates that the inherent intra-domain heterogeneity significantly exacerbates the difficulty of cross-domain alignment. Consequently, a critical research question naturally arises: \textit{How can we construct a universal foundation model for multi-domain heterogeneous graphs without relying on manually defined meta-paths?}
\begin{figure}[t]
\centering
\includegraphics[width=0.95\linewidth]{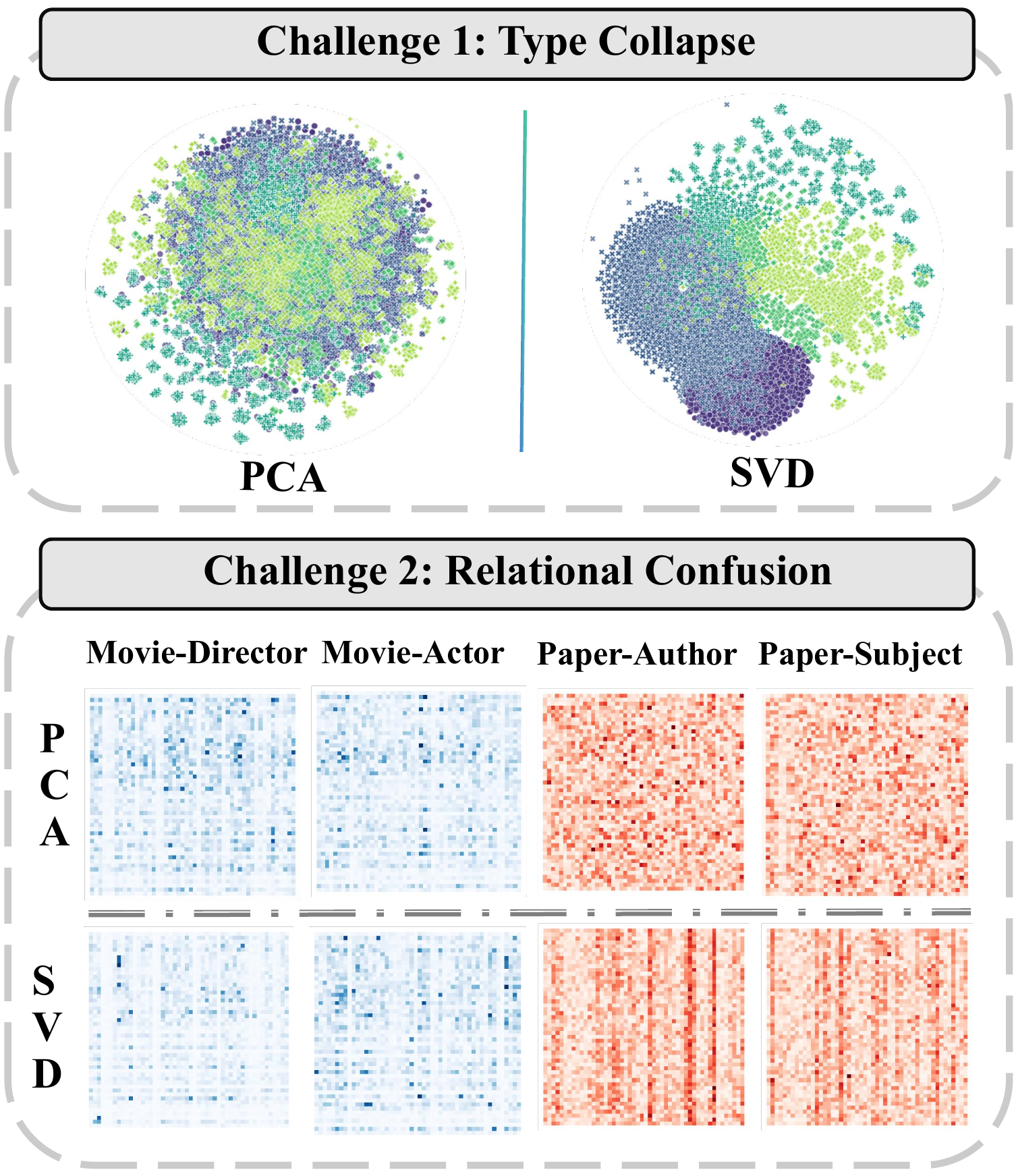}
\caption{Multi-domain heterogeneous graphs face two fundamental challenges. \textbf{Challenge 1} visualizes the node distributions on the ACM and IMDB datasets, where different colors represent node types from different domains. \textbf{Challenge 2} illustrates the relation reconstruction errors after applying traditional feature alignment methods, with color intensity indicating the magnitude of the errors.}
\label{motivation2}
\end{figure}
The core prerequisite for multi-domain pre-training is to overcome the incommensurability of input spaces across domains~\cite{featalign}. To achieve this, existing multi-domain GFMs designed for homogeneous graphs typically employ input-level global alignment techniques, such as Principal Component Analysis (PCA)~\cite{pca} or Singular Value Decomposition (SVD)~\cite{svd}, to project multi-domain node features into a unified dimensional space before pre-training. However, this strategy implicitly assumes that all node features, regardless of their intrinsic types and relation contexts, can be embedded within a shared metric space. This assumption is fundamentally violated in heterogeneous graphs, where different types of nodes possess fundamentally distinct semantic meanings and structural roles~\cite{hgnn,hgt}. To empirically validate this limitation, we conduct an analysis on two heterogeneous datasets: the academic network ACM and the movie network IMDB~\cite{magnn}. Specifically, by employing traditional alignment methods, we analyze the aligned features from these datasets. Visualizations of multi-type node distributions and structural reconstruction errors reveal that: (1)  \underline{Type Collapse}: different node types exhibit pronounced distributional overlap post-alignment, rendering them indistinguishable; and (2) \underline{Relation Confusion}: structural errors increase substantially, indicating that the alignment process disrupts original relation semantics. These phenomena demonstrate that coarse-grained global alignment not only distorts type-specific information but also overlooks the critical role of relation heterogeneity. 

The aforementioned analysis reveals a critical insight: unlike homogeneous graphs that merely need to address cross-domain distribution shifts, node semantics in multi-domain heterogeneous graphs are jointly determined by their attributes and cross-type relations. Relying solely on feature alignment leads to representation space mismatch. This indicates that an effective alignment mechanism must resolve the intra-domain mismatch among multi-type nodes and heterogeneous relations. Consequently, the core challenge lies in tackling a dual-level distribution shift: \ding{182} \textbf{Cross-type Feature Shift}, where different node types possess feature spaces with varying dimensions and distributions; and \ding{183} \textbf{Intra-domain relation Gap}, where the mixed distribution formed by heterogeneous relations exhibits inconsistencies across different subspaces. Therefore, an effective cross-domain alignment mechanism necessitates the simultaneous modeling of feature semantics and relation structures.

To tackle these challenges, we propose a novel plug-and-play relation-driven alignment framework, Decoupled relation Subspace Alignment (DRSA), which fundamentally rethinks the alignment mechanism in cross-domain heterogeneous graphs. Unlike conventional methods that enforce a unified representation within the feature space, DRSA decouples feature semantics from relation structures and achieves alignment at the level of a relation subspace by explicitly modeling cross-type interactions. Specifically, we introduce a dual-relation subspace projection mechanism that factors cross-type interactions into a shared low-rank subspace and captures relation dependencies via bilinear projections, thereby ensuring semantic consistency of relations within a unified latent space. In addition, we propose a feature–structure decoupled mechanism, where aligned features are decomposed into a semantic projection component and a structural residual term. This design enables the model to preserve feature consistency while adaptively capturing intra-domain variations induced by relation structures. To optimize the model, we adopt an alternating minimization strategy~\cite{bcd} , which progressively aligns relation structures and refines feature representations through a two-stage procedure. As each subproblem is convex and admits a closed-form solution, the overall optimization enjoys stable convergence.


The main contributions of this work are summarized as follows:
\begin{itemize}
    \item We systematically analyze the structural origins of the \emph{negative transfer} phenomenon in multi-domain heterogeneous graph pre-training, and reveal the fundamental limitations of traditional global feature alignment methods, which inevitably induce type collapse and relation confusion when handling multi-type nodes.

    \item We propose a plug-and-play relationly-driven alignment framework, DRSA. By decoupling feature semantics from network topology and explicitly coordinating cross-type interactions within a relation subspace, DRSA effectively avoids the distortion of type-specific semantics. To the best of our knowledge, \textbf{this is the first exploratory work that eliminates meta-path dependency and addresses text-free multi-domain heterogeneous graph foundation models}.

    \item Extensive experiments on multiple real-world benchmark datasets demonstrate that DRSA can be seamlessly integrated as a universal preprocessing module, consistently and significantly improving the cross-domain and few-shot knowledge transfer capabilities of state-of-the-art graph foundation models.
\end{itemize}



\section{Related Work}
\label{sec:related}

\subsection{Heterogeneous Graph Self-Supervised Learning}
Heterogeneous graph self-supervised learning can be categorized into metapath-based and free-metapath methods. Metapath-based methods transform into homogeneous graphs by predefined metapaths. DMGI~\cite{dmgi} and HDMI~\cite{hdmi} learn node consistency by maximizing mutual information between node-level and graph-level representations. HeCo~\cite{heco} introduces both network schema and meta-path views for contrastive learning, while HGMAE~\cite{hgmae} reconstructs metapath-based edges and node features from a generative perspective. MUG~\cite{mug} achieves unified cross-domain input by leveraging a dimension-aware module. These methods are primarily designed for single-domain settings and heavily depend on manually designed metapaths. In multi-domain scenarios, each domain may involve multiple metapath graphs, resulting in substantial computational overhead during pre-training.

Meta-path-free methods aim to remove reliance on handcrafted priors. SHGP~\cite{shgp} introduces structural clustering-based pseudo-labeling in heterogeneous graphs. HERO~\cite{hero} employs a self-expression matrix to capture homophily in heterogeneous graphs, while SCHOOL~\cite{school} optimizes the affinity matrix via spectral clustering focusing on homophilic information. RMR~\cite{rmr} introduces preservation, masking, and reconstruction mechanisms into relation subgraphs. RASH~\cite{rash} models both homophily and heterophily from a relation perspective. Prompt-based methods further enhance downstream performance by introducing type prompts and meta-path prompts~\cite{hetgpt,hgprompt}. However, in multi-domain heterogeneous settings, the structural discrepancies across node types and domains are significantly amplified, making it difficult for these methods to achieve effective knowledge sharing and alignment.

\subsection{ Multi-Domain Pre-trained Graph Foundation Models}
Recently, multi-domain graph foundation models have attracted increasing attention. These methods aim to pre-train on multiple source-domain graphs and transfer to unseen target domains to evaluate generalization~\cite{gfm,gfm2}. Existing approaches typically follow a two-stage paradigm: multi-domain pre-training and cross-domain adaptation~\cite{edgeprompt,gpf,gppt}. For example, GCOPE~\cite{gcope} introduces virtual nodes to mitigate structural discrepancies across graphs; MDGPT~\cite{mdgpt} incorporates domain tokens to enhance cross-domain discriminability; SAMGPT~\cite{samgpt} employs structural tokens to align topological information. Furthermore, MDGFM~\cite{mdgfm} and SA2GFM~\cite{sa2gfm} focus on robustness under domain shifts; BRIDGE~\cite{bridge} proposes alignment risk regularization with a mixture-of-experts mechanism; GRAVER~\cite{graver} and RAG-GFM~\cite{RAG-GFM} improve generalization via substructure transfer and retrieval-augmented strategies.

These methods typically perform input-level alignment (e.g., SVD or PCA) prior to pre-training. However, such global alignment implicitly assumes that nodes from different types share a common feature space, which is invalid in heterogeneous graphs. Given the intrinsic semantic disparity across node types, this coarse alignment inevitably induces metric collapse. The resulting distorted input further propagates through the pre-training, ultimately undermining the effectiveness of cross-domain knowledge transfer.

\section{Preliminaries and Problem Statement}
\label{sec:prelim}

\subsection{Heterogeneous Graph}
A heterogeneous graph is formally defined as $\mathcal{G} = (\mathcal{V}, \mathcal{E}, \mathcal{X}, \phi, \psi)$, where $\mathcal{V}$ and $\mathcal{E}$ represent the sets of nodes and edges, respectively. $\mathcal{X} = \{\mathbf{X}^{(t)}\}_{t \in \mathcal{T}}$ is the collection of feature matrices associated with each node type. The mapping functions $\phi: \mathcal{V} \to \mathcal{T}$ and $\psi: \mathcal{E} \to \mathcal{R}$ assign each node and edge to its corresponding type, where $\mathcal{T}$ and $\mathcal{R}$ denote the sets of node and edge types, satisfying $|\mathcal{T}| + |\mathcal{R}| > 2$. For each relation $r \in \mathcal{R}$, the topological structure is represented by an adjacency matrix $\mathbf{A}^{(r)} \in \{0,1\}^{n_{src} \times n_{dst}}$.

\subsection{Problem Formulation}
Given a collection of $K$ unlabeled source-domain heterogeneous graphs $\mathcal{G}_{s} = \{\mathcal{G}_1, \mathcal{G}_2, \cdots, \mathcal{G}_K\}$, where each graph $\mathcal{G}_i$ originates from a distinct domain $\mathcal{D}_i$, our objective is to pre-train a universal graph encoder $f_\theta$ that generalizes to an unseen target-domain heterogeneous graph. Given that graph data from different domains follow distinct feature distributions, the core prerequisite for multi-domain graph pre-training is achieving alignment of node features.

Therefore, before applying the pre-training encoder $f_\theta$, it is imperative to construct a structure-preserving alignment mapping. For each domain $\mathcal{D}_i$ and node type $\tau \in \mathcal{T}$, we aim to learn an alignment operator $\text{Aligner}(\cdot)$ that projects the raw features $\mathbf{X}_{\tau}$ into a synchronized latent manifold $\mathbf{H}_{\tau} \in \mathbb{R}^{n_\tau \times d}$ (where $d$ is the unified embedding dimension), while preserving the heterogeneous topological constraints:
\begin{equation}
    \mathbf{H}_{\tau} = \text{Aligner} \Big( \mathbf{X}_{\tau}, \{\mathbf{R}_r\}_{r \in \mathcal{R}} \Big)
\end{equation}
Subsequently, the universal graph foundation model $f_\theta$ is optimized over the aligned graphs.


\begin{equation}
    \min_{\theta} \sum_{i=1}^K \mathcal{L}_{pre}(\hat{\mathcal{G}}_i, f_\theta)
\end{equation}
where $\hat{\mathcal{G}}_i$ represents the graph with aligned embeddings $\mathbf{H}_i$. In this work, we remain agnostic to the specific choice of pre-training tasks $\mathcal{L}_{pre}$, as our primary goal is to demonstrate that an effective decoupled relation alignment serves as a fundamental prerequisite for the transferability of graph foundation models.

\begin{figure*}[t]
\centering
\includegraphics[width=0.98\textwidth]{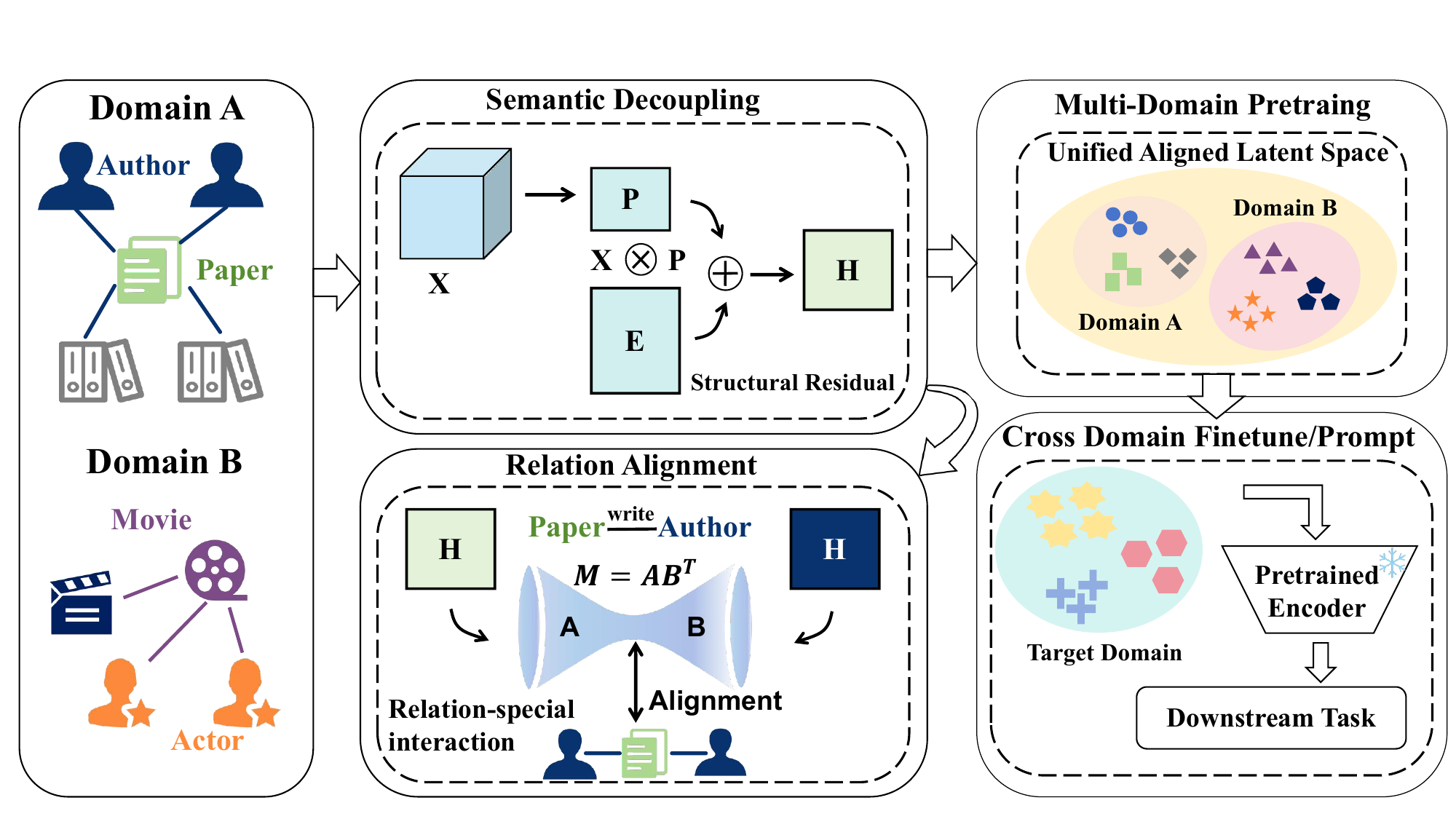}
\caption{Overview of the DRSA framework. DRSA decouples node features into semantic projections and structural residuals, aligns cross-type relations via a low-rank bilinear operator, and learns a unified latent space for multi-domain pretraining and downstream adaptation. It is a plug-and-play alignment module compatible with existing graph foundation models.}
\label{model}
\end{figure*}

\section{Methodology}
In this section, we propose Decoupled relation Subspace Alignment (DRSA) to learn a unified latent space for multi-domain heterogeneous graphs by jointly modeling feature semantics and relation structures. Instead of enforcing direct global feature alignment, DRSA formulates alignment as a structured latent representation learning problem, where cross-type interactions are captured via relation-aware bilinear operators, and node representations are decomposed into a semantic projection component and a structural residual term. This design enables the model to preserve transferable feature information while flexibly adapting to relation-induced variations. The overall objective is optimized through an alternating block coordinate descent procedure, which iteratively refines latent representations and feature projections with closed-form updates, ensuring stable convergence. The framework is shown in the Fig.\ref{model}.

\subsection{Dual-relation Subspace Projection}

In heterogeneous graphs, nodes of different types are associated with type-specific feature spaces, and interactions occur across heterogeneous domains. Relations between node types encode critical structural information. To capture cross-type interactions and achieve alignment of heterogeneous relationships, we introduce a dual-relation subspace projection mechanism. 

Instead of learning an independent, dense transition matrix for every relation type, which is highly prone to overfitting and ignores the shared characteristics of node types across different relations—we factorize the relation operator into type-specific bases. For each node type $\tau \in \mathcal{T}$, we assign two subspace projection matrices: an outgoing projection $\mathbf{A}_\tau \in \mathbb{R}^{k \times \rho}$ and an incoming projection $\mathbf{B}_\tau \in \mathbb{R}^{k \times \rho}$, where $\rho \ll k$. For a connected node pair with source type $\phi_s$ and destination type $\phi_d$, the relation-specific interaction operator $\mathbf{M}_r$ is constructed via a bilinear composition:
\begin{equation}
    \mathbf{M}_r = \mathbf{A}_{\phi_s} \mathbf{B}_{\phi_d}^\top
\end{equation}
Crucially, rather than optimizing $\mathbf{A}_\tau$ and $\mathbf{B}_\tau$ jointly with the node representations, we sample them from a Gaussian distribution ($\mathcal{N}(0, \sigma^2)$) and fix them as non-trainable random projections throughout the alignment process. Fixing $\mathbf{A}_\tau$ and $\mathbf{B}_\tau$ reduces the number of learnable parameters and simplifies the optimization, which helps avoid potential instability caused by the strong coupling between relation operators and latent representations in bilinear formulations. Moreover, random projections have been proven to approximately preserve geometric structures in high-dimensional spaces, providing a diverse set of projection directions~\cite{random_proj_survey, randomFeature, randomproject, ALL-IN}. As a result, the constructed operators $\mathbf{M}_r = \mathbf{A}_{\phi_s} \mathbf{B}_{\phi_d}^\top$ can still capture heterogeneous interactions effectively, without introducing additional trainable parameters.

Given node latent aligned features $\mathbf{H}_{\phi_s}$ and $\mathbf{H}_{\phi_d}$, the relation structure is reconstructed as:
\begin{equation}
\hat{\mathbf{R}}_r = \mathbf{H}_{\phi_s} \mathbf{M}_r \mathbf{H}_{\phi_d}^\top
\end{equation}
where $\mathbf{M}_r$  is a relation-specific bilinear operator  bridging different node types. This formulation enables efficient modeling of heterogeneous interactions within a unified latent space.

\subsection{Decoupled Latent Feature}
The dual-relation subspace projection effectively aligns heterogeneous topological structures; the learned node representations must also preserve the intrinsic semantic information encoded in the original features. However, in real-world heterogeneous graphs, feature semantics and relation topology often reside in inherently different spaces.

A naive formulation that enforces a rigid linear projection $\mathbf{H}\tau = \mathbf{X}\tau \mathbf{P}_\tau$ is therefore overly restrictive, as it implicitly assumes that complex relation dependencies can be fully explained by feature transformations alone. In practice, this leads to an undesirable trade-off between feature fidelity and structural consistency, limiting the expressiveness of the learned representations.

To overcome this limitation and endow the model with the flexibility to capture structure-specific variations, we introduce a feature–structure decoupled representation. Specifically, we decompose the target-aligned feature of each node type into a semantic component and a structural residual:
\begin{equation}
    \mathbf{H}_\tau = \mathbf{X}_\tau \mathbf{P}_\tau + \mathbf{E}_\tau
\label{H=PX+E}
\end{equation}
where $\mathbf{X}_\tau \in \mathbb{R}^{n_\tau \times d_\tau}$ is the raw feature matrix, $\mathbf{P}\tau \in \mathbb{R}^{d_\tau \times k}$ is a type-specific semantic projection operator that maps raw features into the shared latent space, and $\mathbf{E}_\tau \in \mathbb{R}^{n_\tau \times k}$ is is a residual term capturing structure-specific variations from linear projection. This formulation is inspired by decomposition-based representation learning,  where semantic and structural components are explicitly separated.

Instead of jointly optimizing all variables, we adopt a **two-phase strategy:
\begin{itemize}
    \item \textbf{Phase 1:} Estimate a structure-driven aligned feature $\mathbf{H}_\tau$ using only relation information.
    \item \textbf{Phase 2:} Project features into the learned structural subspace and absorb realtion inconsistencies into the residual term.
\end{itemize}
This decomposition prevents feature signals from interfering with structural alignment. The overall objective can be formalized as:
\begin{equation}
\begin{aligned}
     \min_{\{\mathbf{H}, \mathbf{P}, \mathbf{E}\}} & \sum_{r} \|\mathbf{R}_r - \mathbf{H}_{\phi_s} \mathbf{M}_r \mathbf{H}_{\phi_d}^\top\|_F^2 \\
    & \quad + \sum_{\tau} \|\mathbf{H}_\tau - \mathbf{X}_\tau \mathbf{P}_\tau - \mathbf{E}_\tau\|_F^2 \\
    & \quad + \beta \sum_{\tau} \|\mathbf{E}_\tau\|_F^2 
    + \gamma \sum_{\tau} \|\mathbf{P}_\tau\|_F^2
\end{aligned}
\end{equation}
Unlike earlier graph foundation models that solely rely on feature pre-alignment while neglecting heterogeneous nodes and relations within the domain, we further inject the complex multi-relation structural information of heterogeneous graphs into the feature alignment process. This decomposition method provides a flexible alignment mechanism to prevent node type collapse and relation confusion caused by aligning heterogeneous graphs across multiple domains.

\subsection{Optimization}
The proposed unified objective involves multiple coupled variables $\{\mathbf{H}_\tau, \mathbf{P}_\tau, \mathbf{E}_\tau\}$, which makes joint optimization computationally expensive and unstable. Therefore, we adopt a two-stage alternating optimization scheme, which can be interpreted as a Block Coordinate Descent (BCD) method.

\subsubsection{Structure-Driven Target Estimation}
In the first stage, we update the latent aligned feature $\mathbf{H}_\tau$ using only relation information, while fixing all other variables. Specifically, for each node type $\tau \in \mathcal{T}$, we solve the following subproblem:
\begin{equation}
    \min_{\mathbf{H}_\tau}  \sum_{r \in \mathcal{R}} \|\mathbf{R}_r - \mathbf{H}_{\phi_s} \mathbf{M}_r \mathbf{H}_{\phi_d}^\top\|_F^2 + \beta \|\mathbf{H}_\tau\|_F^2
\end{equation}
where the above subproblem denotes the relatoion structural error across all heterogeneous relations connected to node type $\tau$. This is a convex quadratic problem with respect to $\mathbf{H}_\tau$, which leads to the following normal equation:
\begin{equation}
    \mathbf{C}_\tau \mathbf{H}_\tau^\top = \mathbf{B}_\tau
\end{equation}
To rigorously handle directed heterogeneous graphs, we split the relation aggregation into outgoing edges ($t \xrightarrow{r} t'$) and incoming edges ($t' \xrightarrow{r} t$). Let $\Sigma_{t'} = \mathbf{H}_{t'}^\top \mathbf{H}_{t'}$ denote the second-order statistics of the neighbor representations. The operators are explicitly defined as:

Let $\mathbf{\Sigma}_\phi = \mathbf{H}_\phi^\top \mathbf{H}_\phi$ denote the second-order statistics. The operators are analytically defined as: The matrices $\mathbf{C}_\tau$ and $\mathbf{B_\tau}$ are defined as:
\begin{align}
    \mathbf{C}_\tau &= \beta \mathbf{I}_k +  \sum_{\tau \xrightarrow{r} \phi_d} \mathbf{M}_r \mathbf{\Sigma}_{\phi_d} \mathbf{M}_r^\top + \sum_{\phi_s \xrightarrow{r} \tau} \mathbf{M}_r^\top \mathbf{\Sigma}_{\phi_s} \mathbf{M}_r \\
    \mathbf{B}_\tau &=  \sum_{\tau \xrightarrow{r} \phi_d} (\mathbf{R}_r \mathbf{H}_{\phi_d} \mathbf{M}_r^\top)^\top +  \sum_{\phi_s \xrightarrow{r} \tau} (\mathbf{R}_r^\top \mathbf{H}_{\phi_s} \mathbf{M}_r)^\top
\end{align}
The updated representation is obtained as $\mathbf{H}_\tau = (\mathbf{C}_\tau^{-1} \mathbf{B}_\tau)^\top$. This stage is purely structure-driven and does not involve feature-dependent terms. It enforces consistency across heterogeneous relations and constructs a unified structural embedding space.

\subsubsection{Feature  Decomposition}
Given the updated latent aligned feature  $\mathbf{H}_\tau$, we update the feature projection and residual by solving:
\begin{equation}
    \min_{\mathbf{P}_\tau, \mathbf{E}_\tau} \|\mathbf{H}_\tau - \mathbf{X}_\tau \mathbf{P}_\tau - \mathbf{E}_\tau\|_F^2 + \beta \|\mathbf{E}_\tau\|_F^2 + \gamma \|\mathbf{P}_\tau\|_F^2
\end{equation}
This subproblem admits closed-form analytical solutions. We first update the semantic projection matrix via Ridge Regression, and then compute the structural residual:
\begin{align}
    \mathbf{P}_\tau &= (\mathbf{X}_\tau^\top \mathbf{X}_\tau + \gamma \mathbf{I})^{-1} \mathbf{X}_\tau^\top (\mathbf{H}_\tau - \mathbf{E}_\tau) \\
    \mathbf{E}_\tau &= \frac{1}{1+\beta} (\mathbf{H}_\tau - \mathbf{X}_\tau \mathbf{P}_\tau)
\end{align}
This stage ensures that the learned feature remains consistent with the feature space while preserving structural variations. The residual term $\mathbf{E}_\tau$ provides the flexibility to capture domain-specific variations and relation topological noise that cannot be fully explained by a linear feature projection. Based on Equation \ref{H=PX+E}, we further update the latent alignment features $\mathbf{H}_\tau$ for different node types from a feature perspective.

The above two stages are performed iteratively until convergence. This procedure can be interpreted as a Block Coordinate Descent method. Although the overall problem is non-convex, each subproblem is convex with respect to its variables, ensuring that the objective value monotonically decreases during optimization.This design integrates structural information from different relation during the feature alignment process while maintaining compatibility across feature spaces of different node types. The overall procedure is summarized in Algorithm \ref{alg:drsa}.

\begin{algorithm}[t]
\caption{Decoupled relation Subspace Alignment (DRSA)}
\label{alg:drsa}
\begin{algorithmic}[1]

\REQUIRE Features $\{X_\tau\}$, Relations $\{R_r\}$, Hyperparams $\{k, \gamma, \beta\}$, Iterations $\{T\}$
\ENSURE Aligned representations $\{H_\tau\}$

\STATE Initialize $H_\tau, P_\tau, E_\tau$
\STATE Initialize projections $A_\tau, B_\tau$

\FOR{$iter = 1$ to $T$}

    \STATE \textbf{// Stage 1: Structure-driven update}
    \FOR{each node type $\tau$}
    
        \STATE Compute $\Sigma_{\tau'} = H_{\tau'}^\top H_{\tau'}$
        \STATE $C_\tau \leftarrow \beta I$, \quad $B_\tau \leftarrow 0$
        
        \FOR{each relation $r: \tau \rightarrow \tau'$}
            \STATE $M_r \leftarrow A_\tau B_{\tau'}^\top$
            \STATE $C_\tau \leftarrow C_\tau + M_r \Sigma_{\tau'} M_r^\top$
            \STATE $B_\tau \leftarrow B_\tau + (R_r H_{\tau'} M_r^\top)^\top$
        \ENDFOR
        
        \FOR{each relation $r: \tau' \rightarrow \tau$}
            \STATE $M_r \leftarrow A_{\tau'} B_\tau^\top$
            \STATE $C_\tau \leftarrow C_\tau + M_r^\top \Sigma_{\tau'} M_r$
            \STATE $B_\tau \leftarrow B_\tau + (R_r^\top H_{\tau'} M_r)^\top$
        \ENDFOR
        
        \STATE $H_\tau \leftarrow (C_\tau^{-1} B_\tau)^\top$
        
    \ENDFOR

    \STATE \textbf{// Stage 2: Feature decomposition}
    \FOR{each node type $\tau$}
    
        \STATE $P_\tau \leftarrow (X_\tau^\top X_\tau + \gamma I)^{-1} X_\tau^\top (H_\tau - E_\tau)$
        \STATE $E_\tau \leftarrow \frac{1}{1+\beta}(H_\tau - X_\tau P_\tau)$
        \STATE $H_\tau \leftarrow X_\tau P_\tau + E_\tau$
        
    \ENDFOR

\ENDFOR

\RETURN $\{H_\tau\}$

\end{algorithmic}
\end{algorithm}

\section{Experiment}
In this section, we conduct extensive experiments to evaluate the effectiveness and transferability of our proposed DRSA framework. We compare DRSA, integrated with various cutting-edge graph foundation models, against a comprehensive set of baselines on challenging multi-domain heterogeneous graph benchmarks.

\begin{table}[h]
\renewcommand{\arraystretch}{1.0} 
\setlength{\tabcolsep}{5.5pt} 
\centering
\caption{Summary of datasets and their details.}
\begin{tabular}{@{}lcccc@{}}
\toprule
\textbf{Datasets} & \textbf{Node Type} & \textbf{Relation} & \textbf{Target} & \textbf{Classes} \\
\midrule

\multirow{4}{*}{DBLP} 
& Author (A): 4057      & \multirow{4}{*}{\begin{tabular}{@{}c@{}}P-A: 19645 \\ P-C: 14328 \\ P-T: 85810\end{tabular}} & \multirow{4}{*}{Author} & \multirow{4}{*}{4} \\
& Paper (P): 14328      & & & \\
& Conference (C): 20    & & & \\
& Term (T): 7723        & & & \\

\midrule

\multirow{3}{*}{ACM} 
& Paper (P): 4019       & \multirow{3}{*}{\begin{tabular}{@{}c@{}}P-A: 13407 \\ P-S: 4019\end{tabular}} & \multirow{3}{*}{Paper} & \multirow{3}{*}{3} \\
& Author (A): 7167      & & & \\
& Subject (S): 60       & & & \\

\midrule

\multirow{3}{*}{IMDB} 
& Movie (M): 4278       & \multirow{3}{*}{\begin{tabular}{@{}c@{}}M-D: 4278 \\ M-A: 12828\end{tabular}} & \multirow{3}{*}{Movie} & \multirow{3}{*}{3} \\
& Director (D): 2081    & & & \\
& Actor (A): 5257       & & & \\

\midrule
\multirow{3}{*}{Aminer} 
& Paper (P): 6564       & \multirow{3}{*}{\begin{tabular}{@{}c@{}}P-A: 18007 \\ P-R: 58831\end{tabular}} & \multirow{3}{*}{Paper} & \multirow{3}{*}{4} \\
& Author (A): 13329     & & & \\
& Reference (R): 35890  & & & \\
\midrule

\multirow{4}{*}{YELP} 
& Business (B): 2614    & \multirow{4}{*}{\begin{tabular}{@{}c@{}}B-U: 30383 \\ B-S: 2614 \\ B-L: 2614\end{tabular}} & \multirow{4}{*}{Business} & \multirow{4}{*}{3} \\
& User (U): 1286        & & & \\
& Service (S): 4        & & & \\
& Rating Levels (L): 9  & & & \\

\bottomrule
\end{tabular}
\end{table}

\begin{table*}[t]
\caption{Node classification performance comparison. The results of the base models coupled with our DRSA module are highlighted with a blue background, and the corresponding absolute improvements are presented in the subsequent row with a pink background.}
\resizebox{\textwidth}{!}{
\centering

\setlength{\tabcolsep}{1.2pt} 
\renewcommand{\arraystretch}{1.4}

\definecolor{myblue}{HTML}{E8F2FA} 
\definecolor{myred}{HTML}{FDECEE}  

\newcommand{\g}[1]{+#1\%}
\newcommand{\drop}[1]{-#1\%}
\newcommand{\bd}[1]{\mathbf{#1}}

\begin{tabular}{l *{12}{c}}
\toprule
\multirow{2}{*}{\textbf{Method}} 
& \multicolumn{3}{c}{\textbf{ACM}} 
& \multicolumn{3}{c}{\textbf{DBLP}} 
& \multicolumn{3}{c}{\textbf{AMiner}} 
& \multicolumn{3}{c}{\textbf{IMDB}} \\
\cmidrule(lr){2-4} \cmidrule(lr){5-7} \cmidrule(lr){8-10} \cmidrule(lr){11-13}
& ACC & AUC & F1 & ACC & AUC & F1 & ACC & AUC & F1 & ACC & AUC & F1 \\
\midrule

MP2V
& $32.61_{\pm 0.99}$ & $50.21_{\pm 0.56}$ & $32.61_{\pm 0.99}$ 
& $24.65_{\pm 0.98}$ & $50.01_{\pm 0.45}$ & $23.97_{\pm 0.93}$ 
& $24.30_{\pm 2.57}$ & $50.20_{\pm 0.55}$ & $21.66_{\pm 1.44}$ 
& $33.44_{\pm 0.92}$ & $49.85_{\pm 0.76}$ & $32.80_{\pm 0.73}$ \\
HeCo 
& $62.47_{\pm 12.44}$ & $82.94_{\pm 7.60}$ & $61.05_{\pm 11.63}$ 
& $55.47_{\pm 10.43}$ & $81.46_{\pm 6.84}$ & $53.17_{\pm 10.28}$ 
& $26.73_{\pm 5.77}$ & $51.54_{\pm 2.18}$ & $21.90_{\pm 3.02}$ 
& $33.57_{\pm 2.57}$ & $50.25_{\pm 1.78}$ & $30.61_{\pm 2.92}$ \\
HGMAE 
& $67.22_{\pm 12.96}$ & $89.23_{\pm 5.01}$ & $61.91_{\pm 14.88}$ 
& $63.36_{\pm 9.09}$ & $92.62_{\pm 3.38}$ & $59.09_{\pm 9.99}$ 
& $23.99_{\pm 16.28}$ & $51.46_{\pm 3.84}$ & $13.44_{\pm 6.08}$ 
& $34.58_{\pm 3.89}$ & $54.61_{\pm 3.49}$ & $28.06_{\pm 5.41}$ \\
\midrule

RMR 
& $50.88_{\pm 14.82}$ & $60.75_{\pm 13.29}$ & $47.63_{\pm 14.84}$ 
& $30.03_{\pm 5.07}$ & $55.85_{\pm 5.43}$ & $24.18_{\pm 4.27}$ 
& $25.92_{\pm 4.27}$ & $52.74_{\pm 3.00}$ & $22.15_{\pm 2.89}$ 
& $34.48_{\pm 2.73}$ & $51.65_{\pm 2.64}$ & $32.09_{\pm 3.56}$ \\
\rowcolor{myblue}
\textbf{+DRSA}
& $\bd{65.07}_{\pm 11.33}$ & $\bd{76.26}_{\pm 11.96}$ & $\bd{61.40}_{\pm 11.01}$ 
& $\bd{82.71}_{\pm 9.06}$  & $\bd{96.69}_{\pm 3.22}$  & $\bd{81.97}_{\pm 9.82}$ 
& $\bd{41.20}_{\pm 8.94}$  & $\bd{64.97}_{\pm 5.57}$  & $\bd{37.36}_{\pm 7.23}$ 
& $\bd{36.78}_{\pm 2.52}$  & $\bd{52.15}_{\pm 2.59}$  & $\bd{32.73}_{\pm 4.48}$ \\
\hdashline
\rowcolor{myred}
\textit{\textbf{Impro.}}
& \g{14.19} & \g{15.51} & \g{13.77}
& \g{52.68} & \g{40.84} & \g{57.79}
& \g{15.28} & \g{12.23} & \g{15.21}
& \g{2.30}  & \g{0.50}  & \g{0.64} \\
\midrule

HGPrompt 
& $56.07_{\pm 9.73}$ & $77.63_{\pm 11.17}$ & $55.42_{\pm 10.29}$ 
& $36.67_{\pm 7.02}$ & $63.82_{\pm 6.61}$ & $33.90_{\pm 6.96}$ 
& $37.60_{\pm 8.49}$ & $62.83_{\pm 6.99}$ & $33.67_{\pm 7.29}$ 
& $33.08_{\pm 2.82}$ & $49.99_{\pm 1.13}$ & $27.71_{\pm 3.95}$ \\
\rowcolor{myblue}
\textbf{+DRSA}
& $\bd{70.90}_{\pm 11.51}$ & $\bd{85.14}_{\pm 11.04}$ & $\bd{69.62}_{\pm 11.88}$ 
& $\bd{63.43}_{\pm 11.02}$ & $\bd{85.21}_{\pm 5.88}$  & $\bd{61.31}_{\pm 10.98}$ 
& $\bd{43.33}_{\pm 9.82}$  & $\bd{67.48}_{\pm 8.04}$  & $\bd{38.57}_{\pm 8.22}$ 
& $\bd{34.73}_{\pm 3.55}$  & $\bd{53.25}_{\pm 3.21}$  & $\bd{29.09}_{\pm 4.47}$ \\
\hdashline
\rowcolor{myred}
\textit{\textbf{Impro.}}
& \g{14.83} & \g{7.51} & \g{14.20}
& \g{26.76} & \g{21.39} & \g{27.41}
& \g{5.73}  & \g{4.65}  & \g{4.90}
& \g{1.65}  & \g{3.26}  & \g{1.38} \\
\midrule

HetGPT 
& $55.91_{\pm 8.57}$ & $71.87_{\pm 5.69}$ & $53.98_{\pm 6.84}$ 
& $32.44_{\pm 5.95}$ & $59.37_{\pm 5.45}$ & $26.18_{\pm 5.71}$ 
& $32.39_{\pm 4.99}$ & $56.56_{\pm 3.54}$ & $27.64_{\pm 3.52}$ 
& $34.82_{\pm 3.13}$ & $52.97_{\pm 4.07}$ & $31.72_{\pm 4.35}$ \\
\rowcolor{myblue}
\textbf{+DRSA}
& $\bd{64.83}_{\pm 8.12}$  & $\bd{75.58}_{\pm 6.01}$  & $\bd{62.82}_{\pm 6.54}$ 
& $\bd{75.72}_{\pm 9.62}$  & $\bd{90.27}_{\pm 5.87}$  & $\bd{73.49}_{\pm 10.28}$ 
& $\bd{43.67}_{\pm 15.77}$ & $\bd{56.85}_{\pm 9.15}$  & $\bd{37.63}_{\pm 8.26}$ 
& $\bd{37.94}_{\pm 5.56}$  & $\bd{56.04}_{\pm 5.93}$  & $\bd{34.91}_{\pm 5.36}$ \\
\hdashline
\rowcolor{myred}
\textit{\textbf{Impro.}}
& \g{8.92}  & \g{3.71}  & \g{8.84}
& \g{43.28} & \g{30.90} & \g{47.31}
& \g{11.28} & \g{0.29}  & \g{9.99}
& \g{3.12}  & \g{3.07}  & \g{3.19} \\
\midrule
GCOPE 
& $49.02_{\pm 7.42}$ & $68.08_{\pm 5.79}$ & $46.43_{\pm 6.63}$ 
& $42.83_{\pm 5.50}$ & $68.36_{\pm 4.89}$ & $41.33_{\pm 5.67}$ 
& $33.12_{\pm 6.49}$ & $59.80_{\pm 6.05}$ & $29.86_{\pm 5.00}$ 
& $35.32_{\pm 2.35}$ & $52.90_{\pm 2.48}$ & $32.49_{\pm 2.61}$ \\
\rowcolor{myblue}
\textbf{+DRSA}
& $\bd{80.48}_{\pm 11.26}$ & $\bd{93.74}_{\pm 5.77}$ & $\bd{80.91}_{\pm 10.42}$ 
& $\bd{79.59}_{\pm 7.05}$  & $\bd{93.41}_{\pm 3.40}$ & $\bd{78.03}_{\pm 7.90}$ 
& $\bd{42.59}_{\pm 7.77}$  & $\bd{68.46}_{\pm 5.88}$ & $\bd{39.10}_{\pm 7.03}$ 
& $\bd{38.76}_{\pm 3.97}$  & $\bd{57.10}_{\pm 5.30}$ & $\bd{36.14}_{\pm 4.45}$ \\
\hdashline
\rowcolor{myred}
\textit{\textbf{Impro.}}
& \g{31.46} & \g{25.66} & \g{34.48}
& \g{36.76} & \g{25.05} & \g{36.70}
& \g{9.47}  & \g{8.66}  & \g{9.24}
& \g{3.44}  & \g{4.20}  & \g{3.65} \\
\midrule

MDGPT 
& $47.40_{\pm 5.66}$ & $66.67_{\pm 5.69}$ & $46.61_{\pm 5.58}$ 
& $32.19_{\pm 2.69}$ & $57.59_{\pm 3.32}$ & $31.57_{\pm 3.05}$ 
& $25.14_{\pm 2.81}$ & $50.21_{\pm 0.95}$ & $22.24_{\pm 1.34}$ 
& $33.41_{\pm 3.42}$ & $52.57_{\pm 3.18}$ & $28.03_{\pm 3.66}$ \\
\rowcolor{myblue}
\textbf{+DRSA}
& $\bd{76.99}_{\pm 11.08}$ & $\bd{91.42}_{\pm 5.87}$ & $\bd{77.18}_{\pm 10.39}$ 
& $\bd{77.70}_{\pm 10.10}$ & $\bd{93.12}_{\pm 6.03}$ & $\bd{76.74}_{\pm 10.43}$ 
& $\bd{41.70}_{\pm 6.12}$  & $\bd{67.17}_{\pm 4.85}$ & $\bd{38.04}_{\pm 4.77}$ 
& $\bd{37.99}_{\pm 5.18}$  & $\bd{56.11}_{\pm 6.98}$ & $\bd{35.66}_{\pm 5.91}$ \\
\hdashline
\rowcolor{myred}
\textit{\textbf{Impro.}}
& \g{29.59} & \g{24.75} & \g{30.57}
& \g{45.51} & \g{35.53} & \g{45.17}
& \g{16.56} & \g{16.96} & \g{15.80}
& \g{4.58}  & \g{3.54}  & \g{7.63} \\
\midrule

SAMGPT 
& $46.34_{\pm 4.82}$ & $65.32_{\pm 4.89}$ & $45.57_{\pm 4.66}$ 
& $31.82_{\pm 3.31}$ & $57.72_{\pm 3.82}$ & $31.29_{\pm 3.53}$ 
& $25.16_{\pm 2.05}$ & $50.82_{\pm 1.57}$ & $22.56_{\pm 1.29}$ 
& $35.10_{\pm 2.46}$ & $53.35_{\pm 2.73}$ & $34.24_{\pm 2.73}$ \\
\rowcolor{myblue}
\textbf{+DRSA}
& $\bd{76.34}_{\pm 12.49}$ & $\bd{90.42}_{\pm 5.50}$ & $\bd{76.32}_{\pm 11.58}$ 
& $\bd{80.58}_{\pm 8.29}$  & $\bd{93.92}_{\pm 4.53}$ & $\bd{79.20}_{\pm 9.25}$ 
& $\bd{43.23}_{\pm 9.69}$  & $\bd{68.11}_{\pm 6.51}$ & $\bd{39.44}_{\pm 8.19}$ 
& $\bd{37.20}_{\pm 4.81}$  & $\bd{55.62}_{\pm 6.41}$ & $\bd{35.32}_{\pm 5.00}$ \\
\hdashline
\rowcolor{myred}
\textit{\textbf{Impro.}}
& \g{30.00} & \g{25.10} & \g{30.75}
& \g{48.76} & \g{36.20} & \g{47.91}
& \g{18.07} & \g{17.29} & \g{16.88}
& \g{2.10}  & \g{2.27}  & \g{1.08} \\

\bottomrule
\end{tabular}
}
\label{tab:node_results_style}
\end{table*}
\subsection{Experimental Setup}

\para{Datasets. }
To evaluate the model’s performance, we used six publicly available benchmark datasets: DBLP, ACM, IMDB, Aminer, Freebase, Yelp, DBLP~\cite{dblp} and ACM~\cite{magnn}, Aminer is a heterogeneous graph dataset of academic papers, IMDB~\cite{magnn} comes from a movie dataset, and Yelp~\cite{yelp} is a commercial heterogeneous graph dataset.

\para{Baselines. }
To comprehensively validate the superiority of DRSA, we benchmark it against ten representative state-of-the-art models, categorized by their distinct pre-training methodologies: \textbf{Metapath-based Methods}: Metapath2Vec(MP2V)~\cite{metapath2vec}, HeCo~\cite{heco}, and HGMAE~\cite{hgmae}. These models address heterogeneity through meta-path-guided neighborhood aggregation. \textbf{Metapath-free Methods}: RMR~\cite{rmr}, HetGPT~\cite{hetgpt} and HGPrompt~\cite{hgprompt}. These approaches explicitly account for the multiple node types and edge types in heterogeneous graphs without requiring conversion to homogeneous graphs. \textbf{Graph Foundation Models}: GCOPE~\cite{gcope}, MDGPT~\cite{mdgpt}, and SAMGPT~\cite{samgpt}. This category encompasses the latest research explorations in multi-domain pre-training for homogeneous graph structures.

Since DRSA is designed as a plug-and-play input alignment module, we integrate it with representative models from the latter two categories, resulting in variants such as RMR + DRSA and GCOPE + DRSA. This setup allows us to isolate and quantify the contribution of DRSA.

\para{Evaluation Protocol. }
We evaluate the model's cross-domain generalization capability to unseen target domains, treating each dataset as a distinct domain. In a leave-one-out fashion, one dataset is designated as the target domain for evaluation, while the remaining datasets are utilized for multi-domain pre-training.

For methods primarily designed for single-domain scenarios, we employ Singular Value Decomposition for feature alignment across different domains, unifying the alignment dimension to 50. To facilitate the learning of universal knowledge across diverse domains, we utilize a parameter-sharing encoder architecture. Specifically, HeCo and HGMAE adopt a HAN~\cite{han} encoder, RMR employs a GAT~\cite{gat} encoder, and all other methods leverage a GCN~\cite{gcn} encoder. The hidden layer dimension for all encoders is consistently set to 256.

In the downstream evaluation, we assess the effectiveness of DRSA on two distinct tasks: few-shot node classification and few-shot link prediction~\cite{fewshot}. For few-shot node classification, $k$ labeled instances per class are sampled for training, with the remainder used for testing. For few-shot link prediction, categories are defined based on different edge types to evaluate whether distinct types of nodes maintain their heterogeneous relation structures. Each $k$-shot setting is repeated 20 times, with different label sampling for each run. We employ three widely-used evaluation metrics for classification tasks: classification accuracy (Acc), average AUC-ROC (AUC), and average F1-score (F1), reporting their means and standard deviations across these runs. All implementations were executed on an NVIDIA RTX 5090 GPU with 32GB memory.

\subsection{Performance Comparison}

\para{Node Classification.} For node classification, we observed that models specifically designed for heterogeneous graphs also exhibit shortcomings in cross-domain scenarios. Meta-path-based methods such as HeCO and HGMAE outperform meta-path-free methods in multi-domain scenarios; meta-path-free methods like RMR and HetGPT, which rely solely on the aggregation of heterogeneous relationships, exhibit more pronounced negative transfer. Table \ref{tab:node_results_style} demonstrates that DRSA can deliver significant performance improvements for such single-domain models. Notably, RMR combined with DRSA achieved a 52.68\% absolute accuracy improvement on the DBLP dataset, while HGPROMPT saw a 13.87\% increase on the ACM dataset.

Recently emerging graph generation models are primarily designed for homogeneous graphs. When directly applied to multi-domain heterogeneous graphs, these methods suffer from severe performance degradation. This is mainly attributed to their reliance on crude global feature alignment techniques. By integrating DRSA as a preprocessing alignment module, DSRA significantly enhances the performance of state-of-the-art graph foundation models, including GCOPE, MDGPT, and SAMGPT, yielding gains of 20\%-30\% across multiple datasets.

This indicates that existing graph foundation models (GFMs) are largely constrained by suboptimal input alignment. By providing a well-calibrated latent space, DRSA effectively enables cross-domain generalization.

\para{Edge Classification.} To evaluate DSRA’s ability to capture heterogeneous relation, we evaluated edge classification performance in Table \ref{tab:edge_results_style}. The results confirm the superiority of our proposed dual-relationship subspace projection mechanism; across all datasets, the DSRA-enhanced models consistently outperformed their baseline versions. Notably, on Yelp—a highly complex and sparse business dataset—SAMGPT+DRSA achieved a significant 21.66\% increase in accuracy. This indicates that DSRA effectively coordinates heterogeneous interactions while aligning node semantics, thereby avoiding the relationship confusion caused by traditional alignment methods.

The results in Tables \ref{tab:node_results_style} and \ref{tab:edge_results_style} demonstrate that DSRA consistently improves the performance of all backbone models across all datasets and evaluation metrics. This confirms that DSRA is a general-purpose and effective input-level alignment module suitable for multi-domain heterogeneous graphs, and it also indicates that the input alignment problem is a primary bottleneck in existing methods.
\begin{table*}[t]
{
\caption{Edge classification performance comparison. The results of the base models coupled with our DRSA module are highlighted with a blue background, and the corresponding absolute improvements are presented in the subsequent row with a pink background.}
\resizebox{\textwidth}{!}{
\centering
\renewcommand{\arraystretch}{1.2}

\definecolor{myblue}{HTML}{E8F2FA}
\definecolor{myred}{HTML}{FDECEE}
\newcommand{\g}[1]{+#1\%}
\newcommand{\drop}[1]{-#1\%}
\newcommand{\bd}[1]{\mathbf{#1}}

\begin{tabular}{l *{9}{c}}
\toprule
\multirow{2}{*}{\textbf{Method}} 
& \multicolumn{3}{c}{\textbf{Yelp}} 
& \multicolumn{3}{c}{\textbf{DBLP}} 
& \multicolumn{3}{c}{\textbf{IMDB}} \\
\cmidrule(lr){2-4} \cmidrule(lr){5-7} \cmidrule(lr){8-10}
& ACC & AUC & F1 & ACC & AUC & F1 & ACC & AUC & F1 \\
\midrule

RMR 
& $37.89_{\pm 9.85}$ & $52.58_{\pm 2.20}$ & $25.79_{\pm 4.06}$ 
& $46.42_{\pm 11.38}$ & $58.28_{\pm 4.92}$ & $34.46_{\pm 5.84}$ 
& $51.77_{\pm 8.27}$ & $51.15_{\pm 1.76}$ & $46.59_{\pm 4.01}$ \\
\rowcolor{myblue}
\textbf{+DRSA}
& $\bd{44.26}_{\pm 7.46}$ & $\bd{56.80}_{\pm 2.83}$ & $\bd{29.78}_{\pm 3.21}$ 
& $\bd{50.84}_{\pm 12.40}$ & $\bd{65.21}_{\pm 6.26}$ & $\bd{38.43}_{\pm 7.74}$ 
& $\bd{57.88}_{\pm 8.16}$ & $\bd{52.64}_{\pm 1.62}$ & $\bd{49.16}_{\pm 3.76}$ \\
\hdashline
\rowcolor{myred}
\textit{\textbf{Impro.}}
& \g{6.37} & \g{4.22} & \g{3.99}
& \g{4.42} & \g{6.93} & \g{3.97}
& \g{6.11} & \g{1.49} & \g{2.57} \\
\midrule

GCOPE 
& $54.50_{\pm 8.67}$ & $70.00_{\pm 3.38}$ & $37.30_{\pm 3.91}$ 
& $33.64_{\pm 13.44}$ & $58.25_{\pm 4.29}$ & $29.87_{\pm 6.96}$ 
& $47.35_{\pm 6.55}$ & $49.81_{\pm 1.45}$ & $44.44_{\pm 4.19}$ \\
\rowcolor{myblue}
\textbf{+DRSA}
& $\bd{81.42}_{\pm 7.70}$ & $\bd{87.42}_{\pm 4.11}$ & $\bd{56.00}_{\pm 4.42}$ 
& $\bd{41.07}_{\pm 12.34}$ & $\bd{61.88}_{\pm 5.17}$ & $\bd{32.38}_{\pm 6.03}$ 
& $\bd{54.10}_{\pm 3.12}$ & $\bd{54.10}_{\pm 3.72}$ & $\bd{48.48}_{\pm 4.44}$ \\
\hdashline
\rowcolor{myred}
\textit{\textbf{Impro.}}
& \g{26.92} & \g{17.42} & \g{18.70}
& \g{7.43}  & \g{3.63} & \g{2.51}
& \g{4.85}  & \g{4.29}  & \g{4.04} \\
\midrule

MDGPT 
& $56.32_{\pm 11.29}$ & $72.09_{\pm 3.77}$ & $38.95_{\pm 5.56}$ 
& $38.59_{\pm 13.64}$ & $53.46_{\pm 2.71}$ & $30.21_{\pm 6.49}$ 
& $47.74_{\pm 8.97}$ & $49.91_{\pm 1.16}$ & $44.15_{\pm 5.30}$ \\
\rowcolor{myblue}
\textbf{+DRSA}
& $\bd{79.29}_{\pm 7.53}$ & $\bd{84.72}_{\pm 3.31}$ & $\bd{54.11}_{\pm 4.55}$ 
& $\bd{42.82}_{\pm 9.12}$ & $\bd{54.05}_{\pm 4.74}$ & $\bd{32.92}_{\pm 4.42}$ 
& $\bd{51.77}_{\pm 7.54}$ & $\bd{54.04}_{\pm 3.39}$ & $\bd{47.90}_{\pm 4.51}$ \\
\hdashline
\rowcolor{myred}
\textit{\textbf{Impro.}}
& \g{22.97} & \g{12.63} & \g{15.16}
& \g{4.23}  & \g{0.59}  & \g{2.71}
& \g{4.03}  & \g{4.13}  & \g{3.75} \\
\midrule

SAMGPT 
& $57.57_{\pm 12.01}$ & $71.58_{\pm 2.85}$ & $38.94_{\pm 5.68}$ 
& $34.41_{\pm 14.15}$ & $53.73_{\pm 3.14}$ & $28.06_{\pm 7.07}$ 
& $48.34_{\pm 7.09}$ & $49.61_{\pm 1.51}$ & $44.97_{\pm 4.66}$ \\
\rowcolor{myblue}
\textbf{+DRSA}
& $\bd{79.23}_{\pm 7.35}$ & $\bd{85.36}_{\pm 3.16}$ & $\bd{53.78}_{\pm 4.15}$ 
& $\bd{44.30}_{\pm 12.14}$ & $\bd{54.38}_{\pm 4.80}$ & $\bd{32.84}_{\pm 5.07}$ 
& $\bd{51.38}_{\pm 7.76}$ & $\bd{53.56}_{\pm 2.63}$ & $\bd{47.55}_{\pm 4.37}$ \\
\hdashline
\rowcolor{myred}
\textit{\textbf{Impro.}}
& \g{21.66} & \g{13.78} & \g{14.84}
& \g{9.89}  & \g{0.65}  & \g{4.78}
& \g{3.04}  & \g{3.95}  & \g{2.58} \\

\bottomrule
\end{tabular}
}
\label{tab:edge_results_style}}
\end{table*}

\begin{figure}[h]
    \centering
    \begin{subfigure}{0.45\linewidth}
        \centering
        \includegraphics[width=\linewidth]{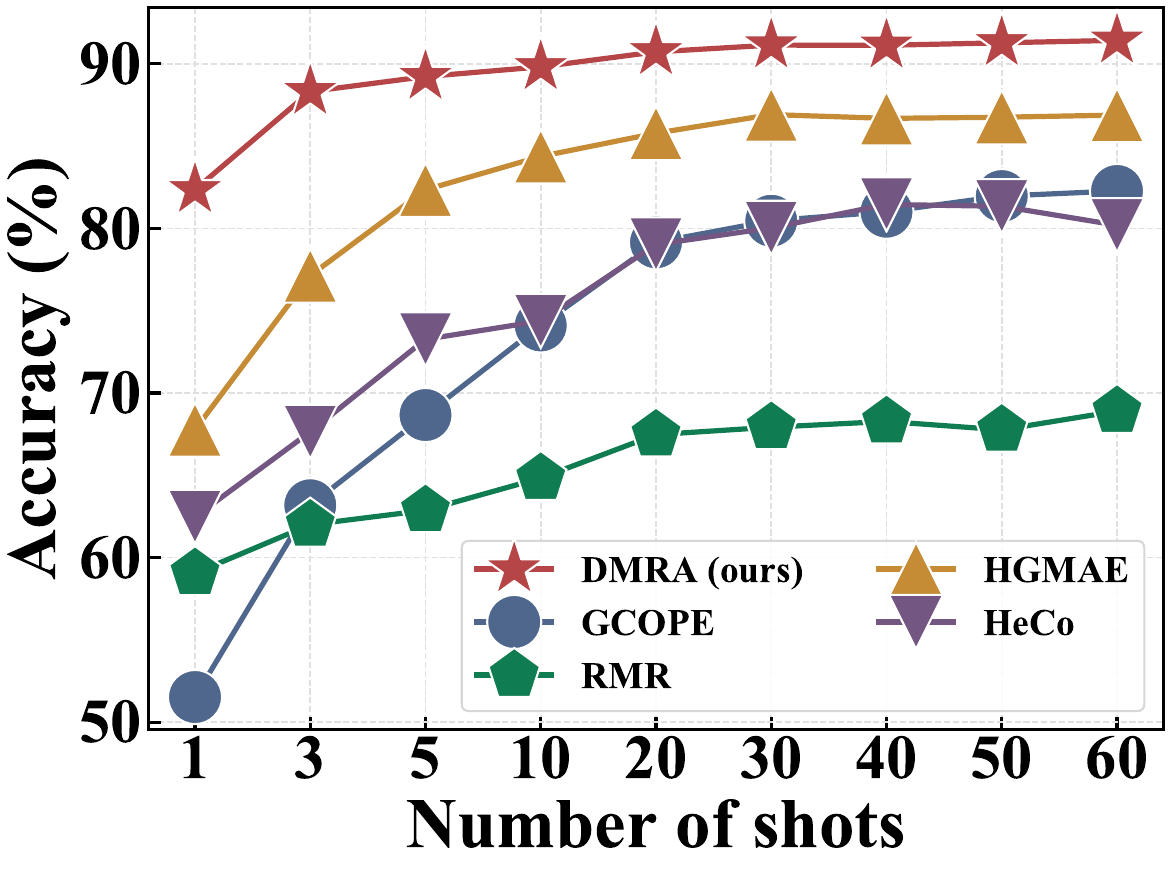}
        \caption{ACM}
    \end{subfigure}
    \hfill
    \begin{subfigure}{0.45\linewidth}
        \centering
        \includegraphics[width=\linewidth]{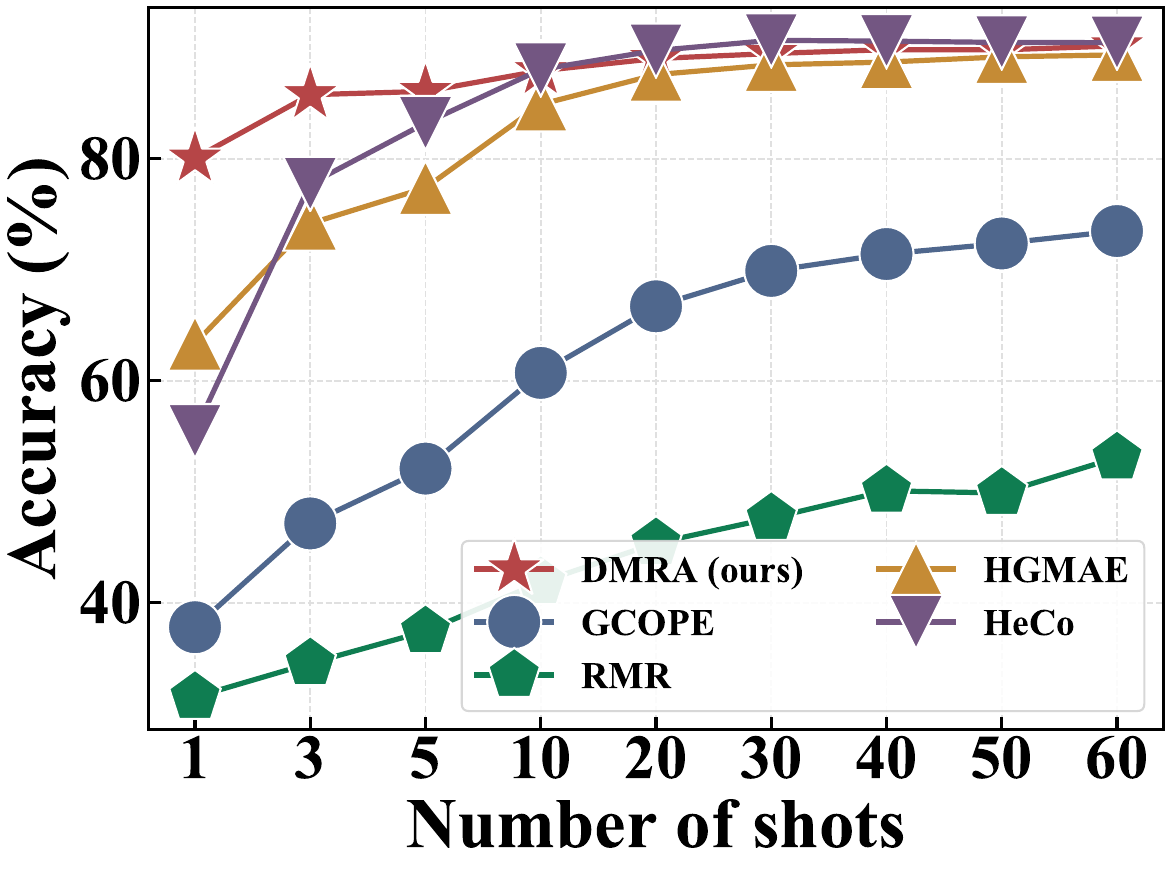}
        \caption{DBLP}
    \end{subfigure}
    
    \vspace{0.2cm}
    
    \begin{subfigure}{0.45\linewidth}
        \centering
        \includegraphics[width=\linewidth]{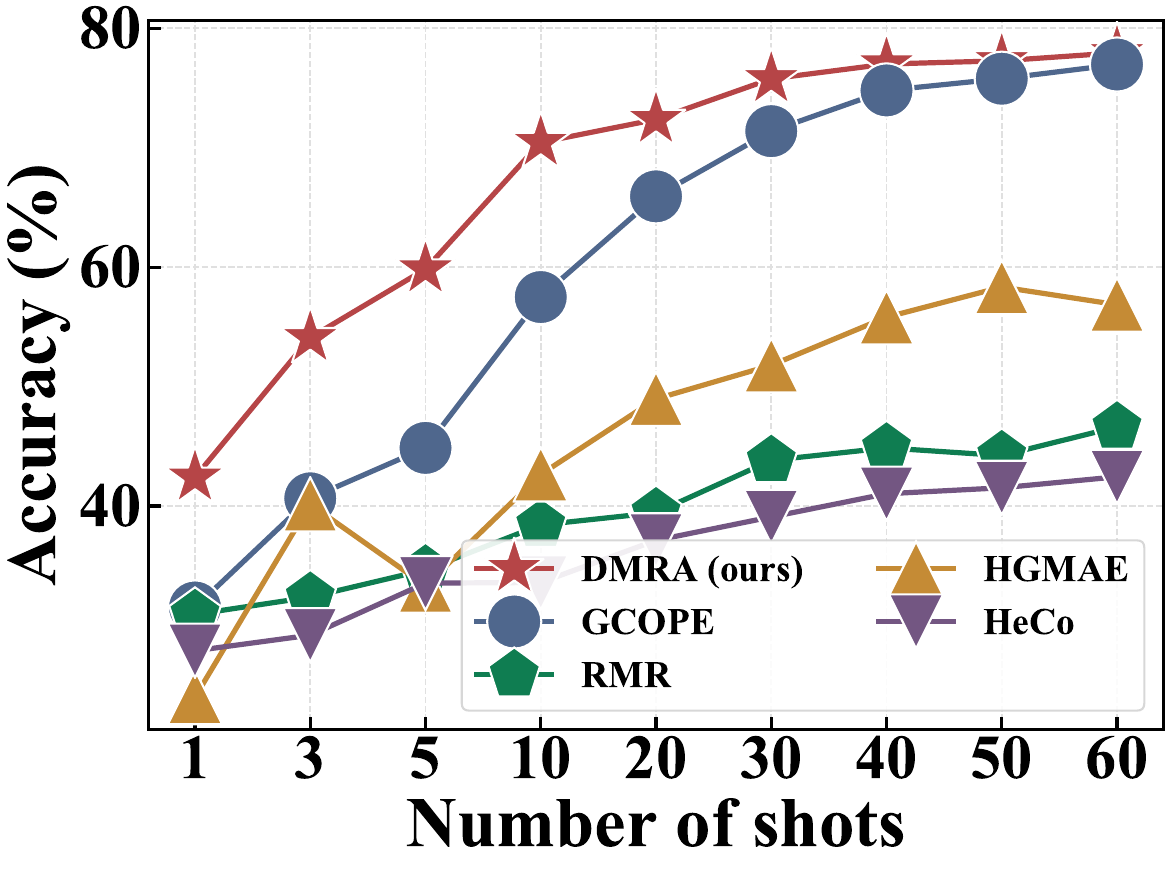}
        \caption{Aminer}
    \end{subfigure}
    \hfill
    \begin{subfigure}{0.45\linewidth}
        \centering
        \includegraphics[width=\linewidth]{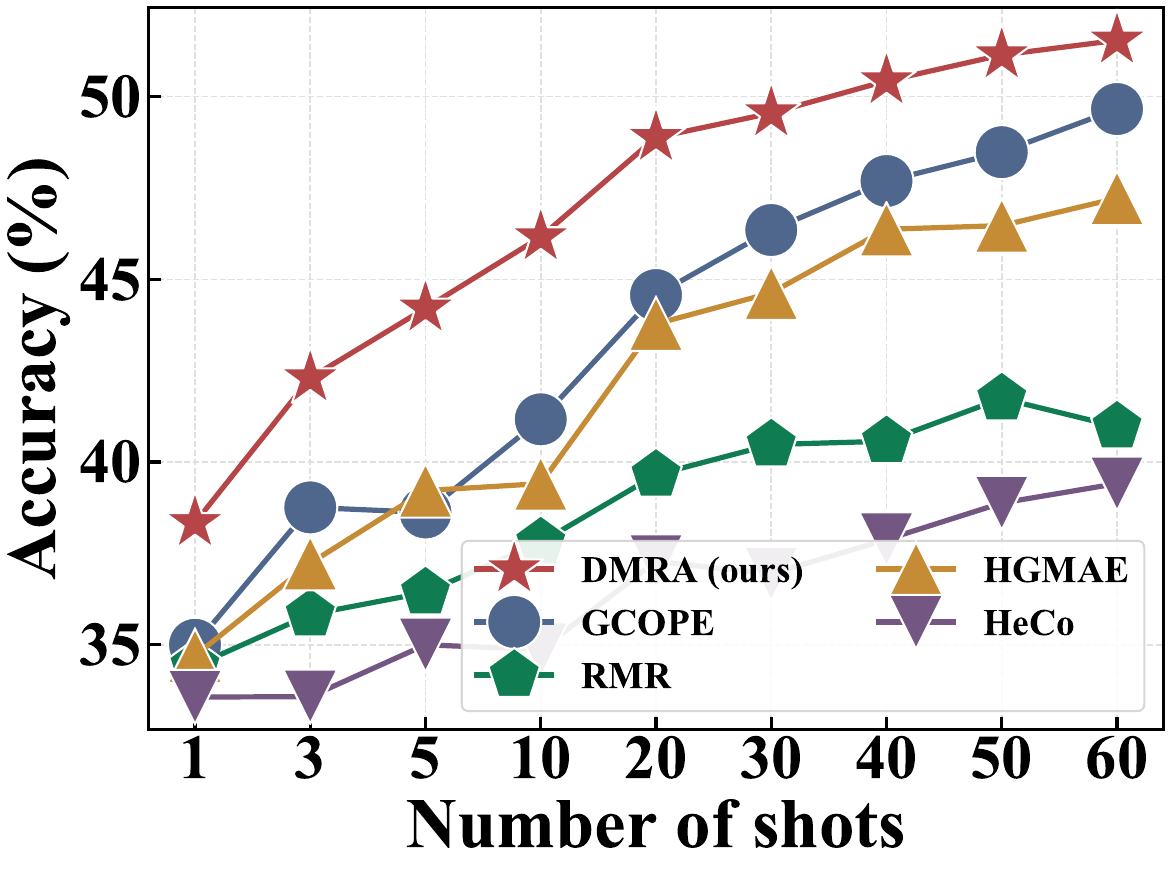}
        \caption{IMDB}
    \end{subfigure}
    
    \caption{ Impact of shots number analysis.}
    \label{fig:few_shot}
\end{figure}
\subsection{Performance on Few-Shot Classification}
To further evaluate the cross-domain generalization capability of DRSA under limited supervision, we conduct few-shot node classification experiments with varying numbers of labeled samples. We adopt GCOPE+DRSA as our representative model. The results are shown in Figure \ref{fig:few_shot}. Overall, DRSA consistently outperforms all baseline methods across all datasets and shot settings. The advantage is particularly pronounced in extremely low-shot ($K \in \{1, 3, 5\}$), where the DRSA-enhanced model demonstrates substantial performance gains. As the number of labeled samples increases, the performance gap between DRSA and the baselines gradually narrows. Nevertheless, DRSA maintains a consistent advantage across all shot settings. In addition, we observe that DRSA exhibits smoother and more stable performance curves compared to baseline methods, suggesting that the learned representations are more robust and less sensitive to sampling variability.

This advantage can be attributed to the input-level alignment mechanism of DRSA. By constructing a unified latent space that preserves both feature semantics and relation consistency, DRSA enables more effective generalization under limited labeled data. These results further validate the effectiveness of the proposed decoupled alignment strategy.

\begin{figure*}[t]
\centering

\begin{subfigure}[t]{0.6\linewidth}
    \centering
    \includegraphics[width=\linewidth]{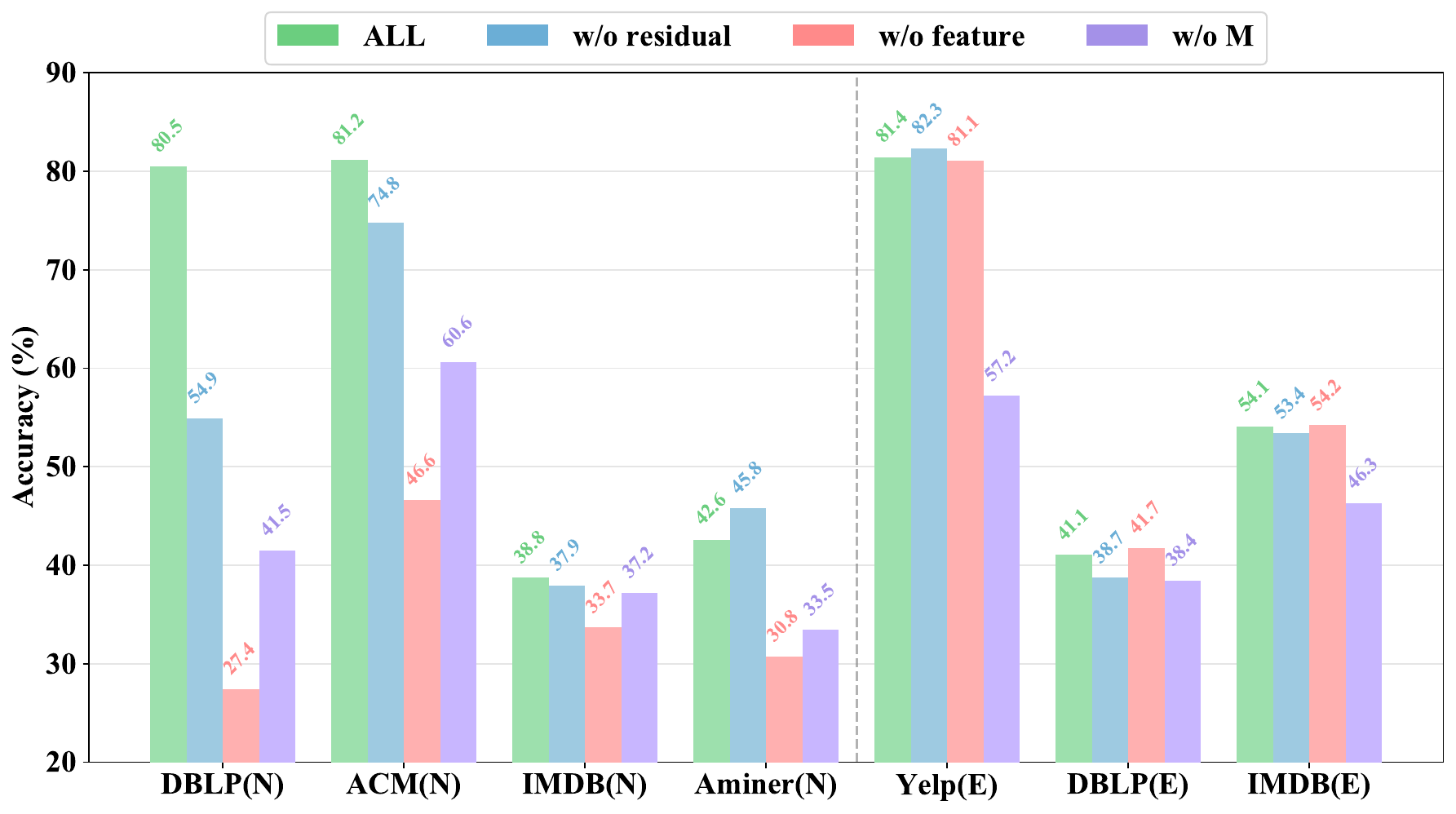}
    \caption{Ablation experiments of different components on four datasets.}
    \label{fig:ablation_main}
\end{subfigure}
\hfill
\begin{subfigure}[t]{0.38\linewidth}
    \centering
    \includegraphics[width=\linewidth]{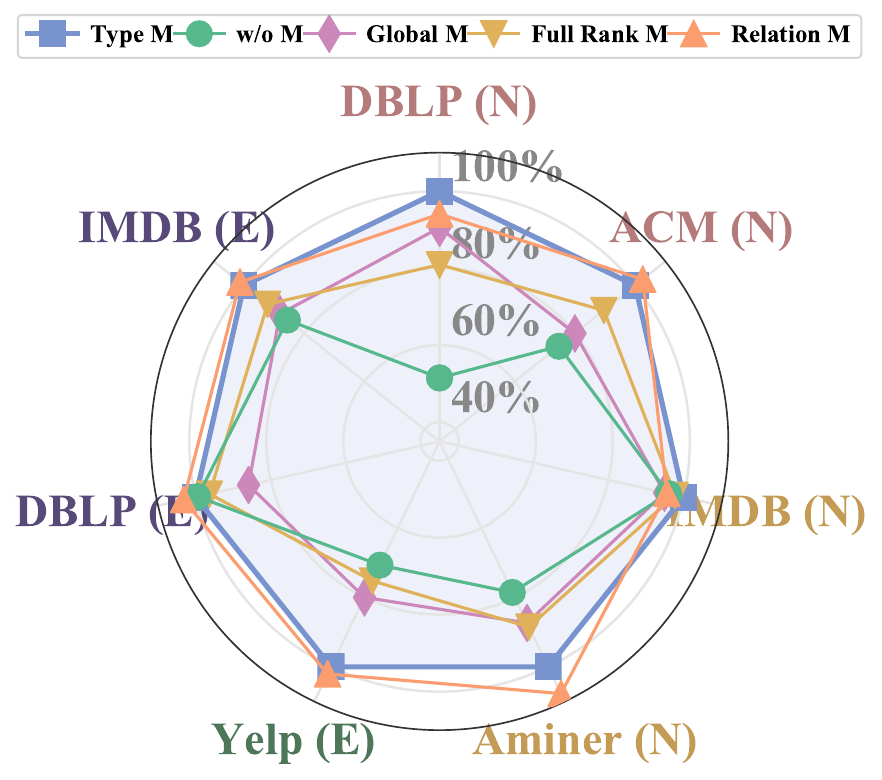}
    \caption{Ablation experiments on variants of different $M$.}
    \label{fig:ablation_M}
\end{subfigure}

\caption{Ablation experiments.}
\label{fig:ablation_combined}
\end{figure*}
\subsection{Ablation Study}
\para{Components Ablation. }To investigate the contributions of individual components within DRSA, we conduct an ablation study, as illustrated in Fig. \ref{fig:ablation_main}. Specifically, we compare the full model (ALL) with three variants: removing the structural residual (w/o residual), removing the feature term (w/o feature), and removing the dual-relation operator (w/o M). (N) represents the node classification task, and (E) represents the edge classification task.

When the structural residual is removed, the aligned feature degenerates to $\mathbf{H}\tau=\mathbf{X}\tau\mathbf{P}_\tau$. This leads to a significant performance drop across all datasets, indicating that $\mathbf{E}$ is essential for capturing structure-specific variations beyond linear feature projection. In heterogeneous graphs, features and relation structures reside in distinct semantic spaces. Enforcing a rigid mapping forces the model to overfit the feature space while neglecting relation inconsistencies.

Removing the feature term also results in noticeable degradation, suggesting that relation structural information alone is insufficient to capture full semantics. Moreover, eliminating feature decomposition disrupts the two-stage optimization mechanism. Structure-driven estimation provides a stable target, while feature projection regularizes the solution and enforces consistency with the input space. This highlights the complementarity between structural and feature information, proving that decoupling their optimization is crucial for achieving stable and effective alignment.

Removing the dual-relation operator causes the most severe degradation. As a relation interaction operator, $\mathbf{M}$ maps features of different node types into a shared relation subspace.  Omitting $\mathbf{M}$ implicitly assumes a unified space across node types, which contradicts the heterogeneous nature of the data and leads to indistinguishable cross-type interactions.

Furthermore, we observe consistent degradation trends in edge classification tasks when ablating these components. Since edge classification fundamentally relies on cross-type interactions, removing the dual-relation operator ($\text{w/o M}$) leads to the most drastic performance drop, as the model fails to map distinct node types into a shared structural space. Similarly, removing the structural residual ($\text{w/o residual}$) noticeably impairs the model's ability to preserve domain-specific topological semantics. the results demonstrate that edge classification is more sensitive to structural modeling, further highlighting the necessity of the dual-relation design.

Overall, the results demonstrate that the three components are complementary and mutually reinforcing: the residual term captures structural variations, the feature decomposition ensures semantic consistency, and the relation operator $\mathbf{M}$ models heterogeneous interactions. Together, they form a unified framework that effectively addresses feature type collapse and relation confusion.

\para{Variants of different $\mathbf{M}$. }
To validate our relation operator design, we compare the node type-based dual projection (Type $\mathbf{M}$) with several variants in Fig. \ref{fig:ablation_M}: the removal operator (w/o $\mathbf{M}$), all relations sharing a single matrix (Global $\mathbf{M}$), applying full-rank projection (Full Rank $\mathbf{M}$), and a parallel low-rank strategy constructed from a relation perspective: $\mathbf{M}=\mathbf{A}_{\psi(s,d)}\mathbf{B}_{\psi(d,s)}^\top$ (Relation $\mathbf{M}$).

Global $\mathbf{M}$ performs poorly due to ignoring relation heterogeneity, while Full Rank $\mathbf{M}$ is prone to overfitting and reduced optimization stability. Type $\mathbf{M}$ and Relation $\mathbf{M}$, on the other hand, exhibit two competitive parallel strategies, each with unique advantages in specific tasks. Specifically, Type $\mathbf{M}$ performs exceptionally well in the node classification task because decomposition using a node type basis better preserves and emphasizes the node-specific semantics during alignment. Conversely, Relation $\mathbf{M}$ demonstrates stronger capabilities in edge classification tasks because its edge-type-centric approach more clearly captures finer-grained topological interaction patterns.

\begin{figure}[h]
    \centering
    \begin{subfigure}{0.45\linewidth}
        \centering
        \includegraphics[width=\linewidth]{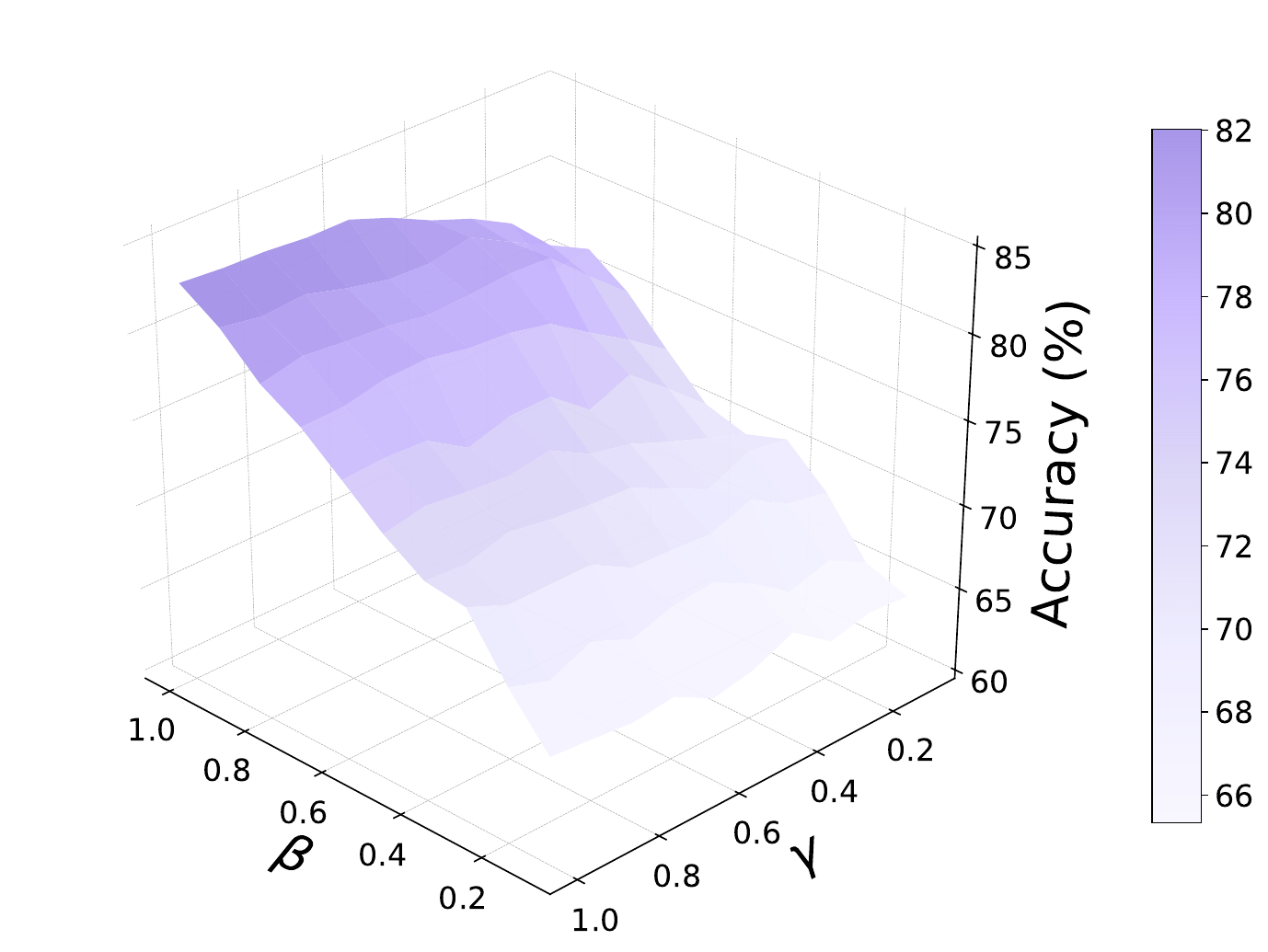}
        \caption{ACM}
    \end{subfigure}
    \hfill
    \begin{subfigure}{0.45\linewidth}
        \centering
        \includegraphics[width=\linewidth]{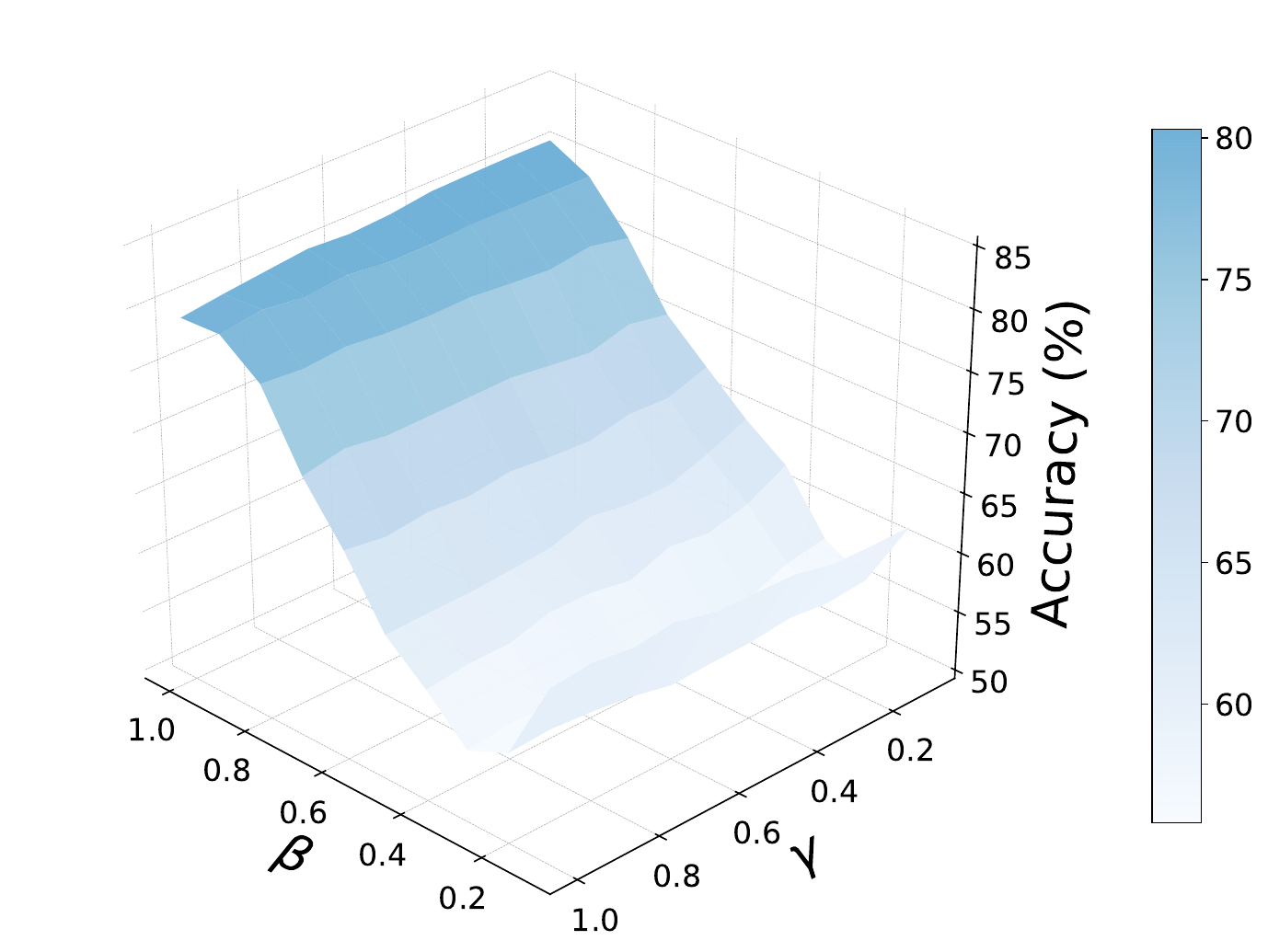}
        \caption{DBLP}
    \end{subfigure}
    
    \vspace{0.2cm}
    
    \begin{subfigure}{0.45\linewidth}
        \centering
        \includegraphics[width=\linewidth]{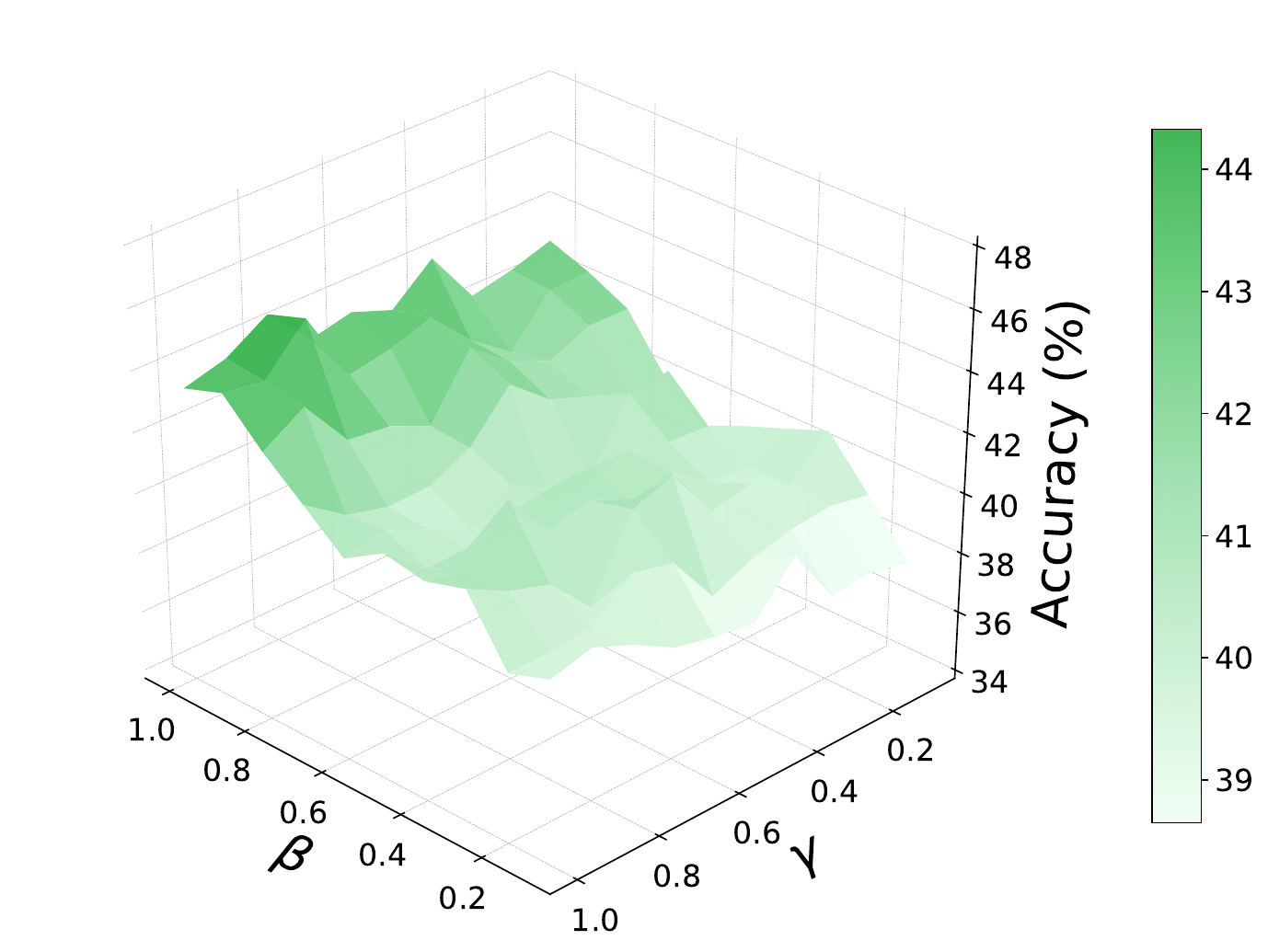}
        \caption{Aminer}
    \end{subfigure}
    \hfill
    \begin{subfigure}{0.45\linewidth}
        \centering
        \includegraphics[width=\linewidth]{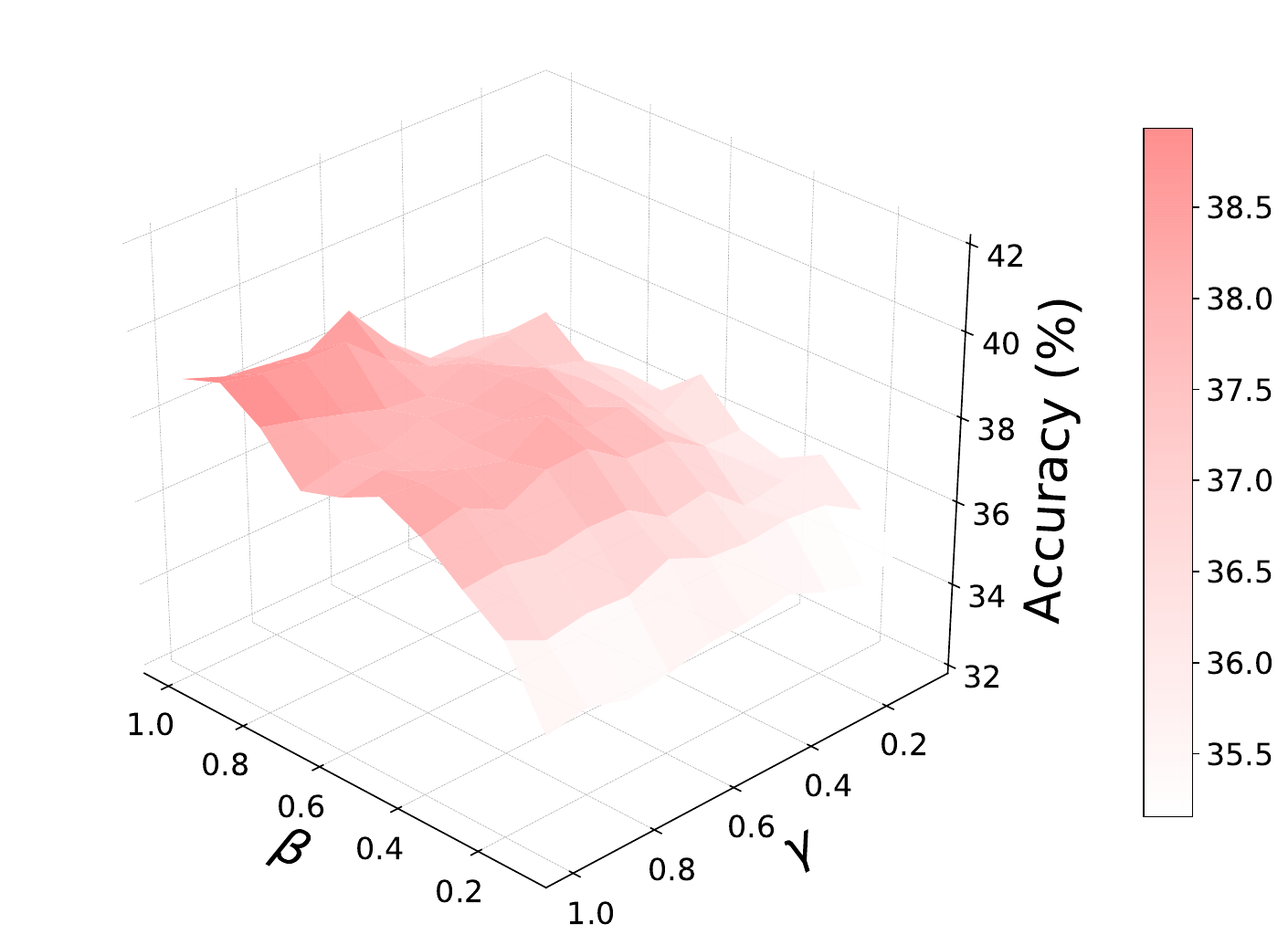}
        \caption{IMDB}
    \end{subfigure}
    
    \caption{Sensitivity of the structural residual penalty $\beta$ and the feature projection regularization $\gamma$ across different datasets.}
    \label{fig:hyperpara}
\end{figure}
\subsection{Hyperparameter Analysis}
In this section, we investigate the sensitivity of the DRSA framework to its core hyperparameters: the structural residual penalty $\beta$ and the feature projection regularization $\gamma$. Figure \ref{fig:hyperpara} illustrates the one-shot classification accuracy across different datasets as $\beta$ and $\gamma$ vary.

\para{Structural Residual Penalty ($\beta$): }The parameter $\beta$ determines the model’s tolerance for structural variations. As shown in Figure \ref{fig:hyperpara}, setting $\beta$ to an extremely small value degrades performance, causing the residual term $\mathbf{E}_\tau$ to dominate during optimization. This results in the model absorbing excessive domain-specific topological noise, thereby weakening the generalization ability of the learned features. Conversely, when $\beta$ is too large, the residual term is excessively suppressed. This forces the model to rely entirely on linear feature projections, causing it to lose the flexibility needed to capture inconsistencies in heterogeneous relation, which confirms that a balanced residual is crucial for decoupling semantic and structural signals.

\para{Feature Regularization ($\gamma$): }The parameter $\gamma$ controls the complexity of the semantic projection matrix $\mathbf{P}_\tau$. Specifically, the performance remains relatively stable under different values of $\gamma$, indicating that the model is not overly sensitive to the strength of feature regularization. This suggests that the feature projection term primarily serves as a stabilizing constraint rather than a dominant factor, making the model less prone to overfitting.

\section{Visualization}
\begin{figure*}[t]
\centering

\begin{minipage}{0.33\linewidth}
    \centering
    \includegraphics[width=\linewidth]{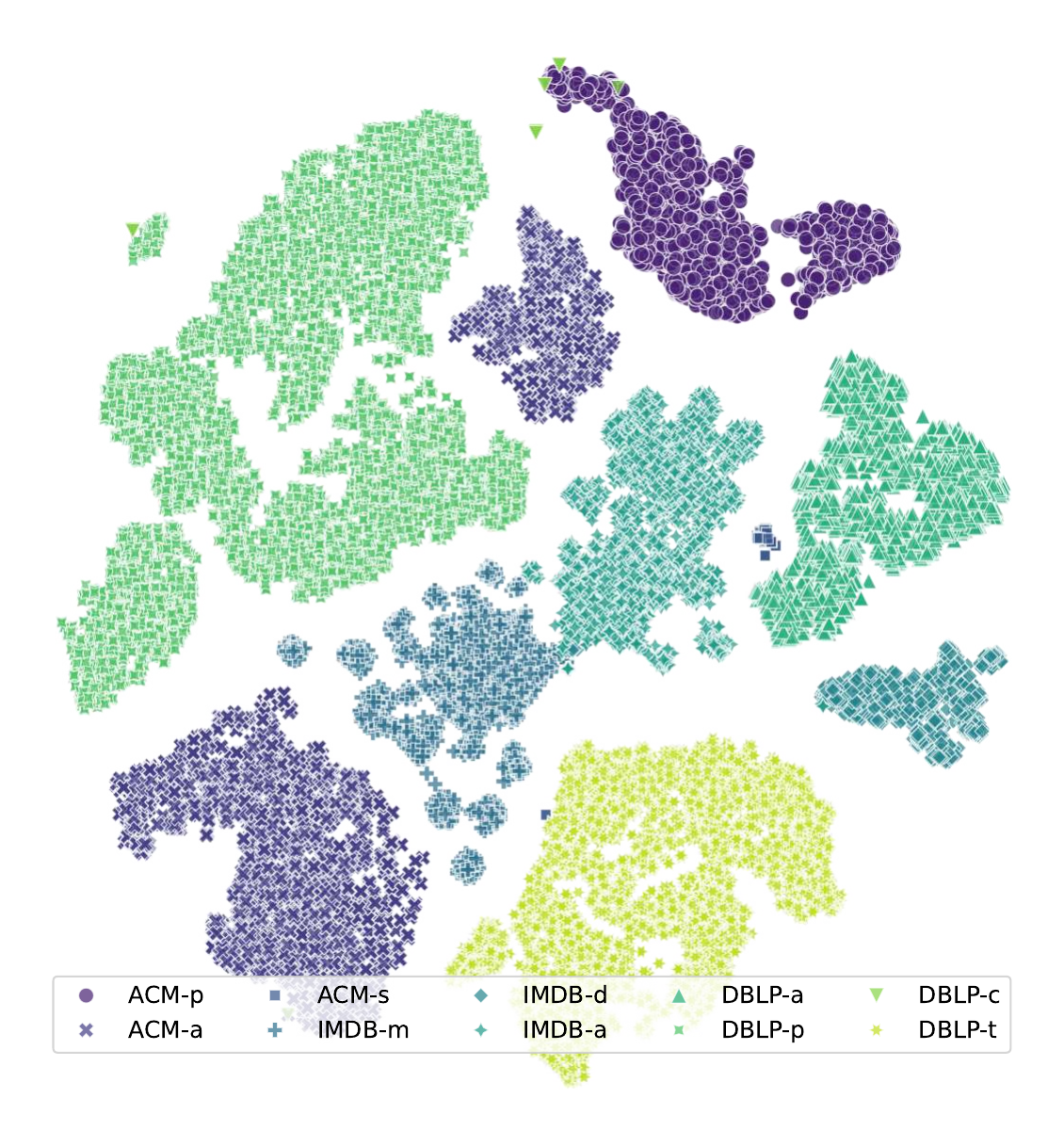}
    \vfill   
    \captionof{figure}{Multi-domain node visualization.}
    \label{fig:vis}
\end{minipage}
\hfill
\begin{minipage}{0.62\linewidth}
    \centering

    \begin{subfigure}{0.23\linewidth}
        \includegraphics[width=\linewidth]{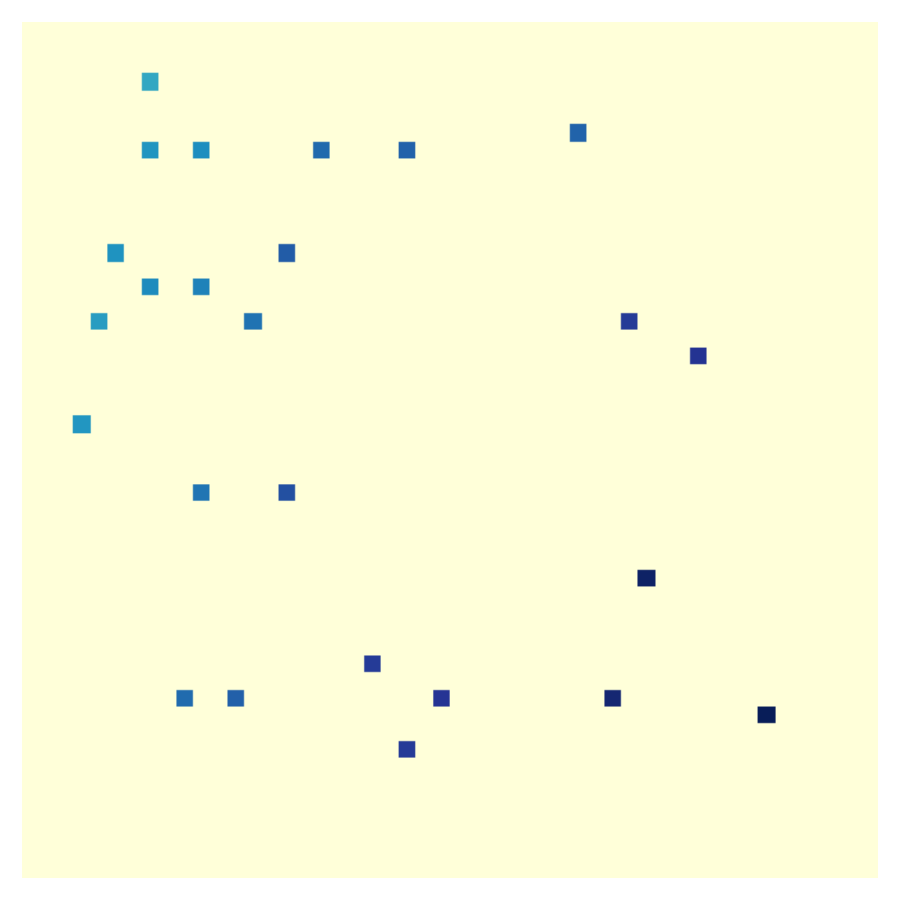}
        \caption{ACM (raw)}
    \end{subfigure}
    \begin{subfigure}{0.23\linewidth}
        \includegraphics[width=\linewidth]{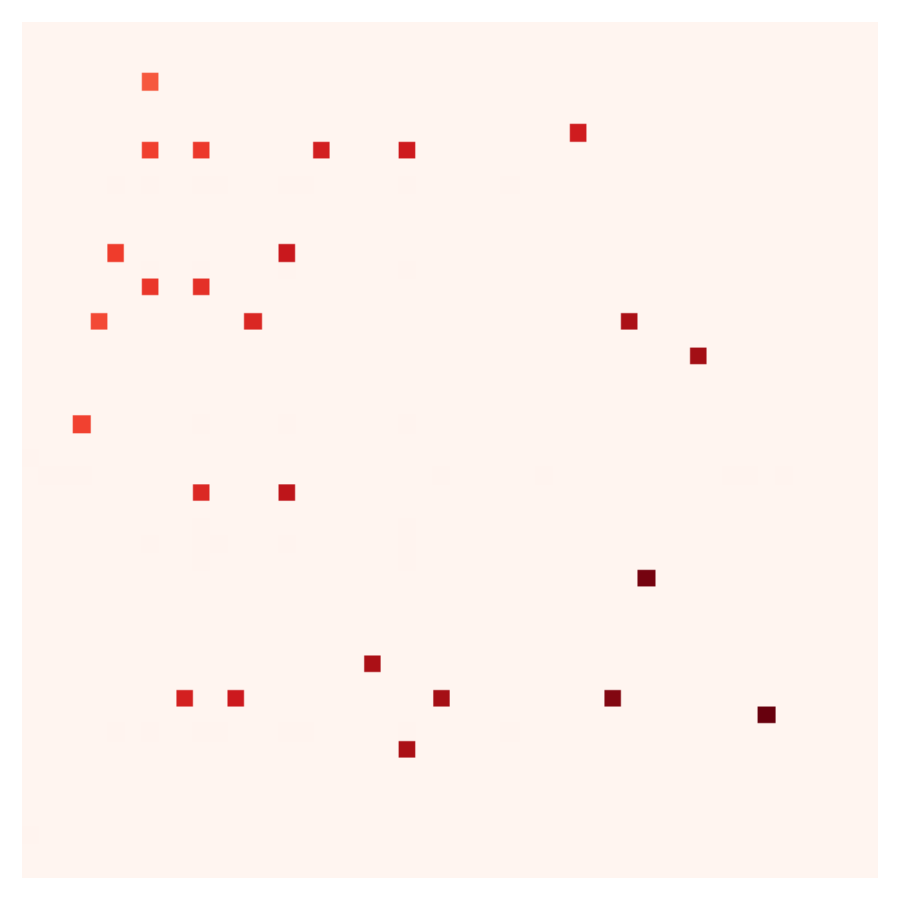}
        \caption{ACM (Ours)}
    \end{subfigure}
    \begin{subfigure}{0.23\linewidth}
        \includegraphics[width=\linewidth]{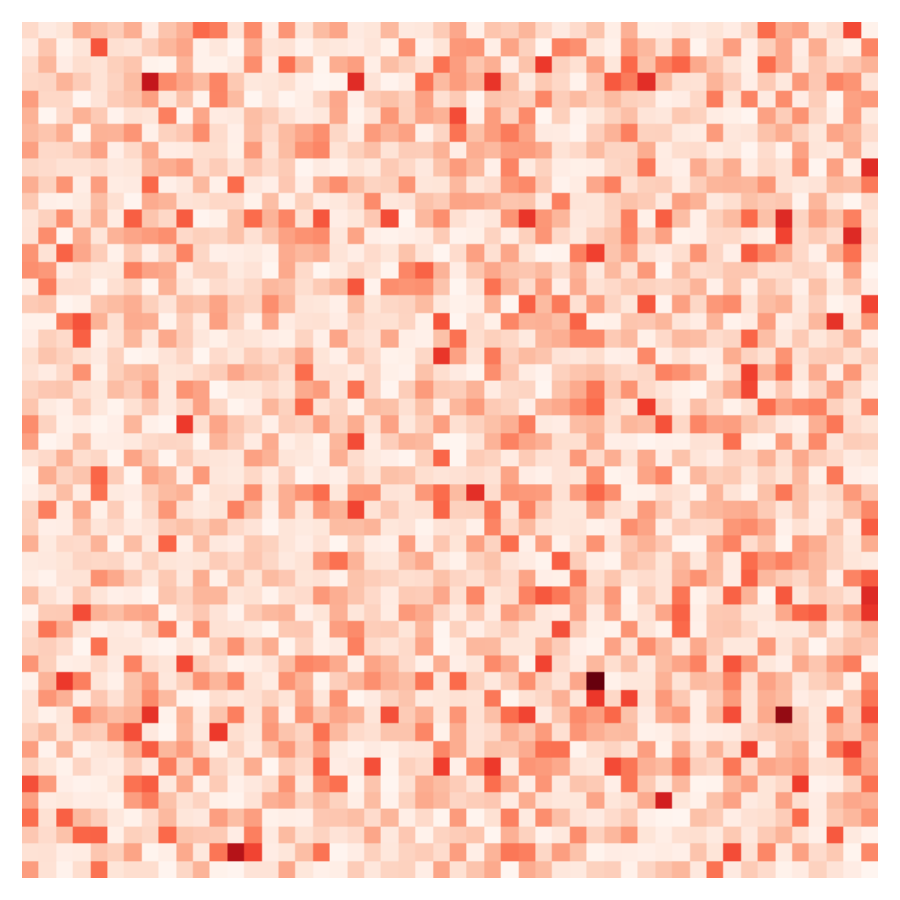}
        \caption{ACM (PCA)}
    \end{subfigure}
    \begin{subfigure}{0.23\linewidth}
        \includegraphics[width=\linewidth]{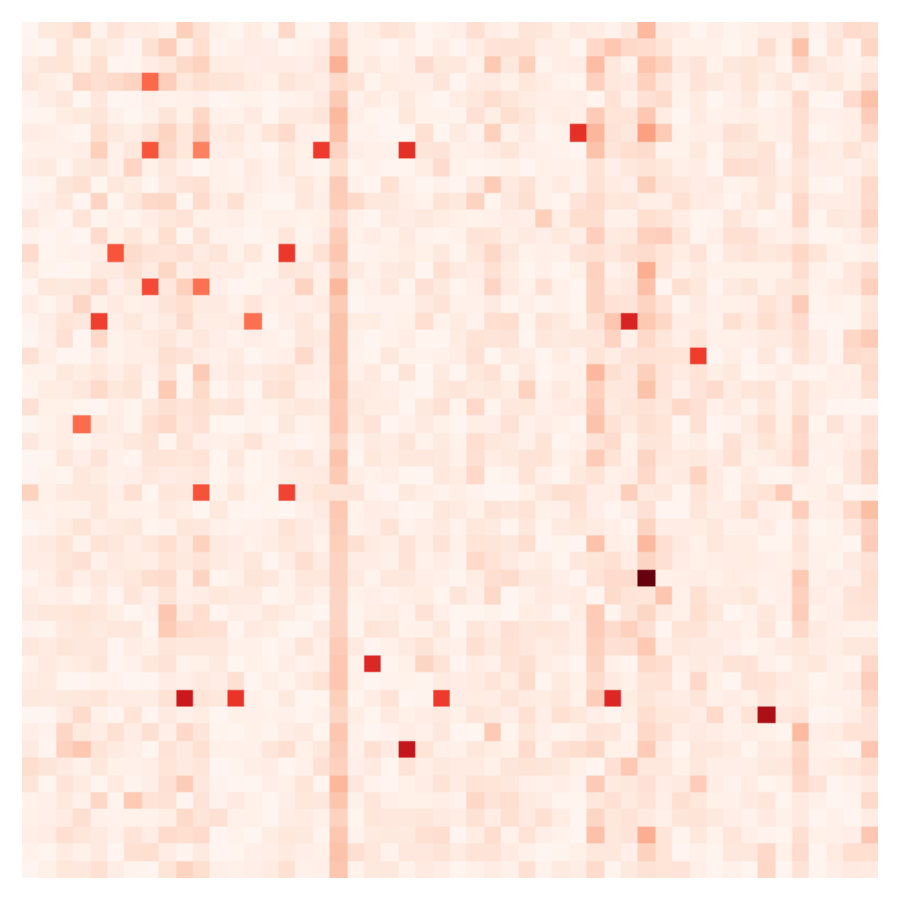}
        \caption{ACM (SVD)}
    \end{subfigure}

    \vspace{0.4em}

    \begin{subfigure}{0.23\linewidth}
        \includegraphics[width=\linewidth]{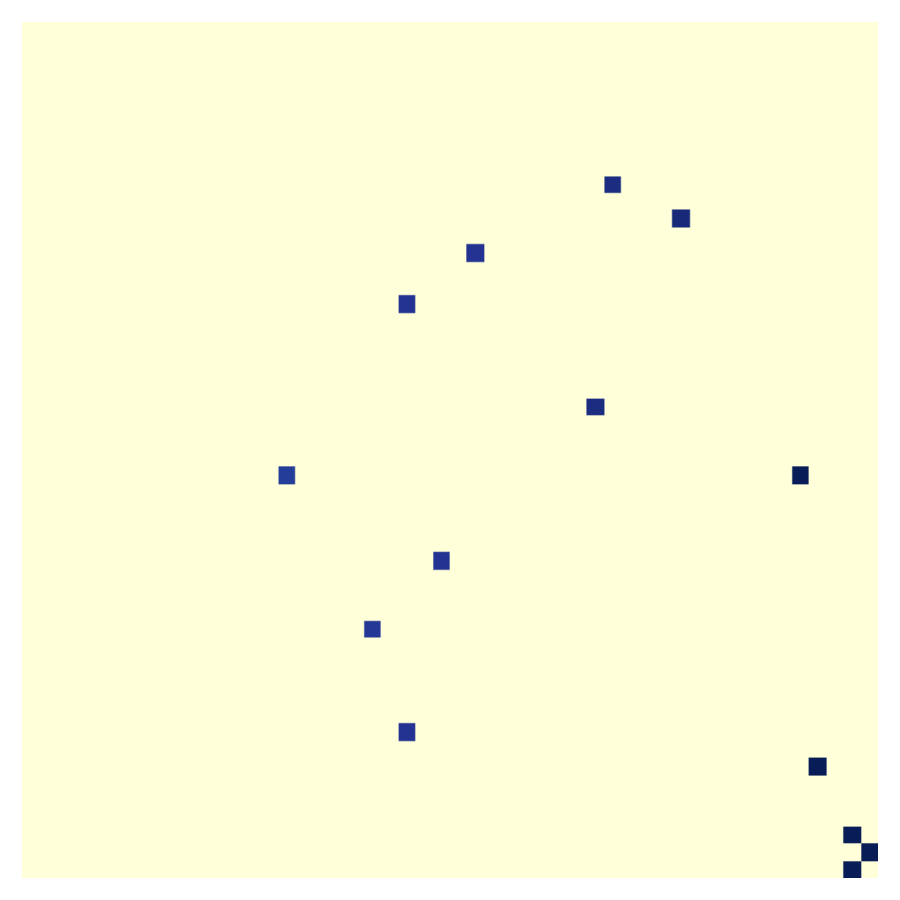}
        \caption{IMDB (raw)}
    \end{subfigure}
    \begin{subfigure}{0.23\linewidth}
        \includegraphics[width=\linewidth]{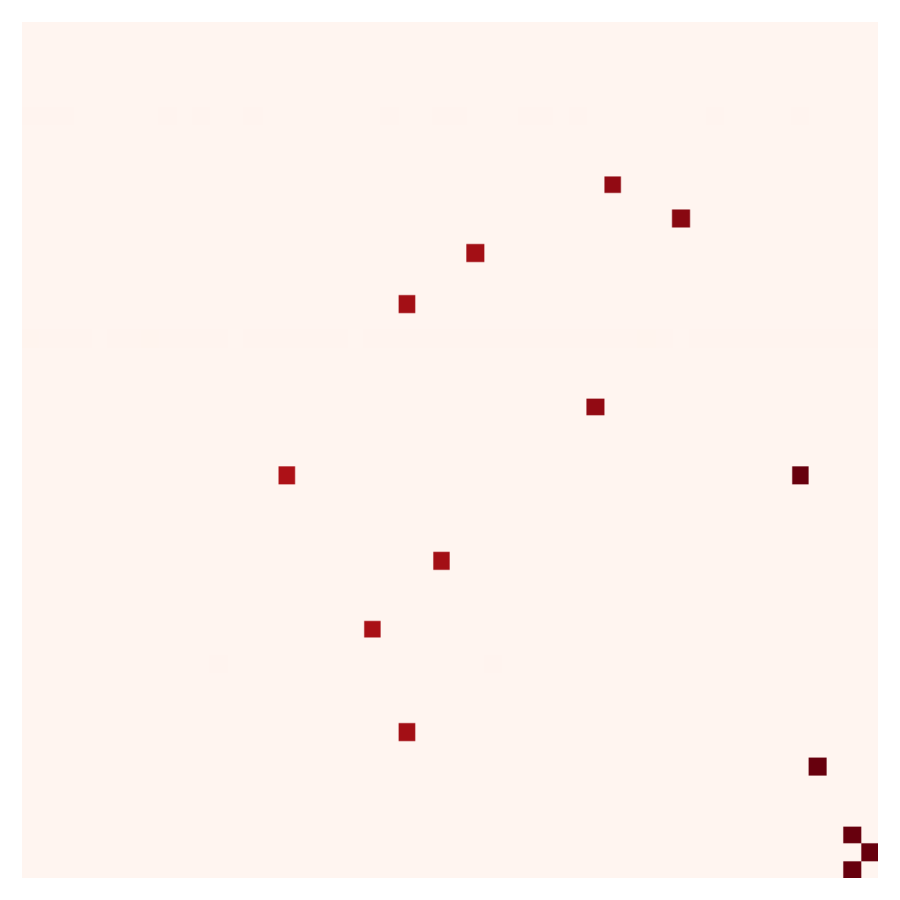}
        \caption{IMDB (Ours)}
    \end{subfigure}
    \begin{subfigure}{0.23\linewidth}
        \includegraphics[width=\linewidth]{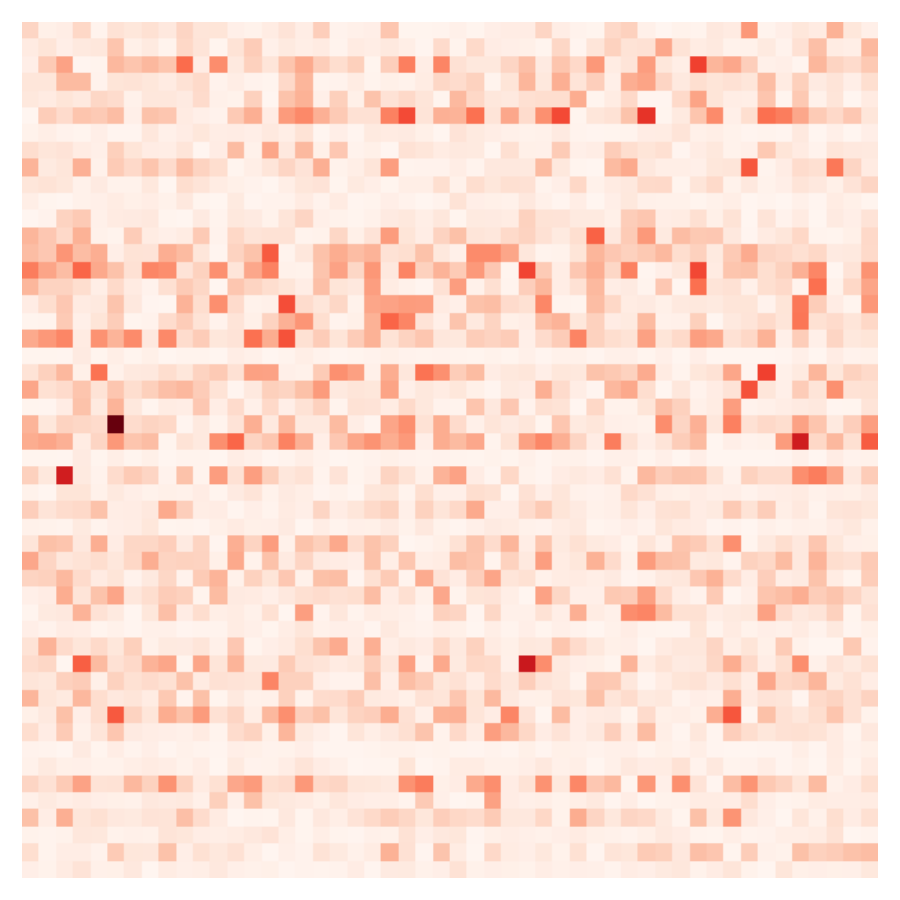}
        \caption{IMDB (PCA)}
    \end{subfigure}
    \begin{subfigure}{0.23\linewidth}
        \includegraphics[width=\linewidth]{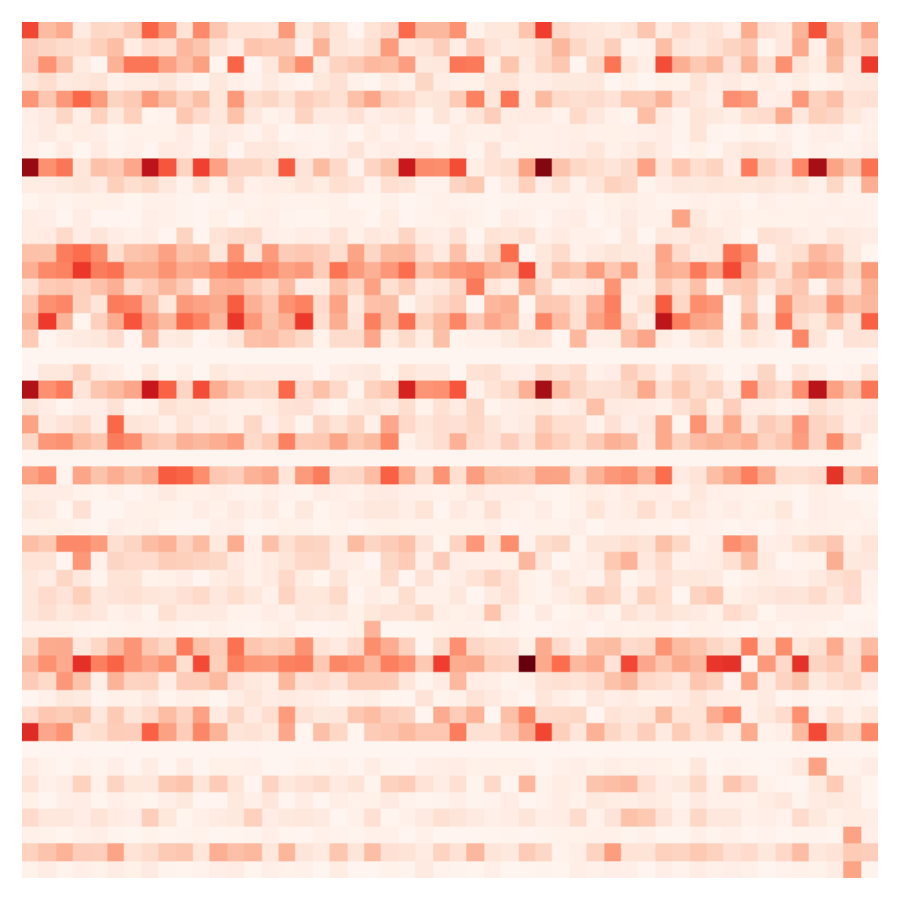}
        \caption{IMDB (SVD)}
    \end{subfigure}
    \vfill   

    \captionof{figure}{Relation structure error on ACM(paper-author) and IMDB(movie-actor).}
    \label{fig:recon}
\end{minipage}

\end{figure*}

To evaluate the aligned features, we visualized node embeddings(Fig. \ref{fig:vis}) and relation reconstructions(Fig. \ref{fig:recon}). Compared to traditional alignment methods, DRSA generates clearly separated clusters across node types and domains, effectively mitigating type collapse. Semantically distinct nodes remain distinguishable within a shared latent space. We further visualized the reconstructed heterogeneous relation patterns. DRSA significantly reduces reconstruction errors, particularly across type relations, and is highly consistent with the true relation space, indicating that the learned latent space maintains relation consistency. These results confirm that DRSA simultaneously maintains semantic distinguishability and structural fidelity, thus avoiding type collapse and relation confusion.

\section{Conclusion}
In this paper, we systematically investigate the "negative transfer" phenomenon in multi-domain heterogeneous graph pre-training and reveal the fundamental limitations of traditional global feature alignment, which inherently induces type collapse and relation confusion when handling multi-type nodes. To address this bottleneck, we introduce Decoupled relation Subspace Alignment (DRSA), a universal, plug-and-play input alignment framework. By innovatively employing a dual-relation subspace projection alongside a feature-structure decoupled representation mechanism, DRSA effectively untangles semantic features from relation topologies. This allows the model to harmonize heterogeneous interactions within a unified latent space without sacrificing type-specific semantic integrity. Extensive empirical evaluations across diverse multi-domain benchmarks confirm that DRSA successfully overcomes dual-level distribution shifts. When integrated with existing state-of-the-art graph foundation models, it delivers substantial and consistent performance gains in both cross-domain generalization and few-shot adaptation tasks. Ultimately, our work provides a highly effective and text-free paradigm for advancing multi-domain heterogeneous graph learning without the reliance on manually crafted meta-paths.

\section*{Acknowledgments}
This work was supported in part by the National Natural Science Foundation of China under Grants 62425605, 62133012, and 62303366, in part by the Key Research and Development Program of Shaanxi under Grants 2025CY-YBXM-041, 2022ZDLGY01-10, and 2024CY2-GJHX-15, and in part by the Fundamental Research Funds for the Central Universities and the Postgraduate Innovation Fund of Xidian University under Grant YJSJ26014.




 
\bibliographystyle{IEEEtran}
\bibliography{refs}

\vfill

\end{document}